\documentstyle[psfig,preprint,eqsecnum,aps]{revtex}

\newcommand{\ie}{{\it i.e.\ }}
\newcommand{\eg}{{\it e.g.\ }}

\newcommand{\expt}[2]{\langle #2 \rangle_{#1}}
\newcommand{\expta}[1]{\langle #1 \rangle}
\newcommand{\dexpt}[1]{\langle\langle #1 \rangle\rangle}
\newcommand{\gint}[2]{\int {[dU] \; {#1} \; e^{#2}}}
\newcommand{\ginta}[1]{\int {[dU] \; e^{#1}}}

\newcommand{\cS}{{\cal S}}

\newcommand{\cW}{{\cal W}}

\newcommand{\Tr}{{\rm Tr}}
\newcommand{\Real}{{\rm Re}}

\begin{document}

\draft                  
\tighten

\preprint{Liverpool Preprint: LTH 365, hep-lat/9608145}

\title{Approximate Actions for Lattice QCD Simulation}

\author{Alan C. Irving}
\address{Theoretical Physics Division, Department of Mathematical Sciences\\
University of Liverpool,
PO Box 147, Liverpool L69 3BX, UK}

\author{James C. Sexton}
\address{School of Mathematics, Trinity College, Dublin 2, Ireland}

\date{August 25, 1996}

\maketitle

\begin{abstract}
We describe a systematic approach to generating approximate actions for
the lattice simulation of QCD.  Three different tuning conditions are
defined to match approximate with true actions, and it is shown that
these three conditions become equivalent when the approximate and true
actions are sufficiently close.  We present a detailed study of
approximate actions in the lattice Schwinger model together with an
exploratory study of full QCD at unphysical parameter values.  We find
that the technicalities of the approximate action approach work quite
well.  However, very delicate tuning is necessary to find an approximate
action which gives good predictions for all physical observables.  Our
best view of the immediate applicability of the methods we describe is
to allow high statistics studies of particular physical observables
after a low statistics full fermion simulation has been used to prepare
the stage.
\end{abstract}


\narrowtext


\section{Introduction}
\label{Sec:Intro}

The design of improved algorithms for simulation with dynamical fermions
has provided an important challenge for lattice QCD.  Perturbative
arguments and numerical experiments suggest that a significant component
of the effect of dynamical quarks can be accounted for by a shift in the
effective lattice spacing.  The quenched approximation takes this view
to its extreme, and ignores all dynamical quark effects except for this
shift in scale.  In most cases, the systematic errors introduced by quenching
have so far turned out to be of the same order as the best
presently achievable statistical errors in full dynamical simulations.
As available computer power grows, and as we study larger
volumes and smaller lattice spacings, we can expect this situation to
change, and to begin to see significant discrepancies between quenched
and full theory results.  Systematic errors due to quenching will then
turn out be larger than full theory statistical errors, and the quenched
approximation will loose its usefulness.

In a recent paper \cite{SW1} it has been shown that the quenched
algorithm is a single member of a large class of approximate QCD
simulation algorithms.  The members of this class interpolate in a
smooth fashion between the extreme cases of quenched QCD simulations on
the one hand, and full dynamical QCD simulations on the other.
Individual members can be considered as different approximations to the
full dynamical QCD action.  So long as the quenched approximation
represents the only significant example of an approximate simulation
algorithm, it is possible to consider it in isolation and to accept its
historically important contributions to lattice QCD while
remaining wary of its mathematical correctness.  However, once one
begins to consider variants which seek to improve on the quenched
approximation without themselves being exact, a number of important
issues arise, and a more critical approach is needed.

First, of course, is the question of what we mean mathematically
by the term \lq approximate action\rq{}.
The qualitative view of quenched QCD which we normally adopt is
that, at fixed lattice spacing and coupling, it produces ensembles
which are reasonably close to the ensembles which would be generated
for full QCD at some shifted coupling.
Reasonable closeness is however a
vague term.  What it tends to mean in practice is that
ensemble expectations of a set of operators in quenched
and full QCD seem to agree within statistical errors, so long
as one allows for an effective lattice spacing shift.

In this paper we consider the general principles which can be applied to
define what is meant by a simulation algorithm based on 
an approximate action in the sense
defined by the quenched approximation.  We will describe a systematic
approach to generating approximate actions and will discuss conditions
which allow the tuning of any candidate approximate action to optimize
its matching to a true action of interest. The methods we develop, while
motivated by the approximate action view, are also applicable
whenever we wish to match results from different lattice simulations.
In particular, the tuning we propose defines the optimum matching
conditions between Wilson and Kogut-Susskind fermion simulations, and between
naive and improved action simulations.

A second issue of concern, when considering simulations
with approximate actions, is the question of their computational 
performance.  The
results reported in \cite{SW1} were obtained on very small systems, and
for these systems it was found that approximate action simulations were
very competitive with exact simulations including full fermions.
However, a question left unanswered was whether this competitiveness
would continue when the method was applied to more realistic situations.
The basic proposal of \cite{SW1} was to attempt an approximation of the
fermion determinant in QCD using Wilson loop operators of increasing
size.  On a small system there are relatively few such operators, but
the number of operators grows very fast as the size of the system increases and the
computational work required would also grow.  Some preliminary data
which we reported in \cite{lat95} for the case of the Schwinger model
seems to bear out this view.  There we found that only very slow
improvements in the quality of the approximation were achieved as Wilson
loops of systematically increasing size were added to the approximate
action.  An independent study of QCD at realistic lattice sizes
\cite{Kilcup} was also somewhat negative.  This study found that the
plaquette operator was actually a very poor approximation to the fermion
determinant in typical large simulation. Both of these results suggest
that very many Wilson loops would be needed to get a numerically useful
approximation in realistic situations.

We report in this paper, however, that the pessimistic view just
described is an artifact of the particular approximation
prescription defined in \cite{SW1}.   The systematic approach
to tuning approximate actions which we develop here allows us
to generate much more aggressive approximations than are otherwise
possible, and we find that it is not necessary to include
Wilson loops of all shapes and sizes to get useful approximate
actions.  Indeed, the best approximate actions in the Schwinger
model and QCD cases which we have studied are found to include
only a few Wilson loop operators of quite surprising
shape and dimension.

An issue which we do not address in this paper is that of
the behaviour of approximate actions in the continuum limit.
The quenched case is clearly problematic in this limit \cite{Bernard}.
However, the strategy we propose here is a numerical one in which
one may choose the most appropriate approximate action
for any given simulation at fixed lattice spacing and coupling.
If one varies the lattice spacing and coupling, then the corresponding
approximate action would also vary.  It seems conceivable to us
that the variations in approximate action which result
would be sufficient to allow a mathematically
correct continuum limit of the approximate action to be taken.

The plan of the remainder of this paper is as follows.
In Section \ref{Sec:CumulantExpansion}, we define a simple cumulant
expansion method which allows direct comparison of the
expectations of operators calculated in two different
path integral measures.  We argue that this expansion method
is the most efficient way of determining the shifts introduced
in expectations of physical observables due to changes in the
underlying path integral measure.

The concept of an approximate action is defined in Section \ref{Sec:Tuning}.
The quenched QCD action is the prototypical example here.   Approximate
actions allow us to define corresponding approximate simulation
algorithms, and we develop in this section three different ways
in which approximate actions might be defined.  We then argue that
all three definitions become equivalent when the approximate action
is sufficiently close to the action which it attempts to approximate.

As an initial test, we consider the Schwinger model. Section
\ref{Sec:SchwingerModel} gives the explicit form of this
two-dimensional model which we have used,
and describes some necessary technical features of
the calculations.  Simulation results are presented in the following
two sections.   In Section \ref{Sec:SchwingerResults},
we consider the quenched approximation in some detail.
Physical quantities calculated here include the static potential
and the mass of the mesonic bound state of the model.  We present
comparisons of these observables calculated in
the quenched approximate simulation and full dynamical simulation.
Section \ref{Sec:Extensions} describes a variety of different
extensions to the quenched approximation.
The analysis begins with the scenario proposed in
\cite{SW1} in which the fermion determinant is expanded
in terms of Wilson loops of increasing size.
We show first that this expansion technique is quite
straightforward
to execute and appears to be numerically stable.
We argue, however,
that this ordered expansion is very slow to converge.
Instead,  we propose
an aggressive approach to developing approximate
actions which directly targets the interesting
physics.  This approach is seen to improve considerably
the slow convergence apparent in \cite{SW1}.

In Section \ref{Sec:QCD}, we describe a trial application of the
approximate action approach to QCD.  The purpose of the
test is to gain some understanding of the potential of the
method for QCD simulation and to identify practical
problems which remain to be solved.  The test
described is for a small lattice with heavy quarks.

Finally, our conclusions are presented in Section \ref{Sec:Conclusions}.


\section{The Comparison of Path Integral Measures}
\label{Sec:CumulantExpansion}

In all lattice simulation, the goal is to calculate expectation values
of interesting operators in some path integral measure.  The measure is
given by some action which we think of as the true action
and denote by $S_t$.
Our goal is to find an action $S_a$ which allows us to approximate the measure
generated by $S_t$.  The prototypical example is given by lattice QCD.
The fundamental degrees of freedom here are the $SU(3)$ link matrices
$U_\mu(x)$ assigned to edges which join nearest neighbouring sites on a
four dimensional hypercube.  A possible `true' action for this system is
the lattice QCD action which includes a Wilson fermion determinant term,
\begin{equation}
  S_t = - \beta \sum_\Box \frac{1}{3} \Real \Tr{U_\Box}
        - \Tr \log M_\kappa^{\dagger}M_\kappa
\end{equation}
while a possible approximate action is the quenched lattice QCD action,
\begin{equation}
 S_a = - \beta' \sum_\Box \frac{1}{3} \Real \Tr{U_\Box}\, .
\end{equation}
$U_\Box$ here represents the product of gauge links around a fundamental
plaquette, while $M_\kappa$ is the Wilson fermion matrix for hopping
parameter $\kappa$.  We have also implicitly allowed for a bare coupling
shift between exact and approximate cases by including different
$\beta$'s in the different actions.

Expectation values of operators $F$ in the path integral
measures generated by $S_t$ and $S_a$ are denoted respectively
by $\expt{t}{F}$ and $\expt{a}{F}$, and are given by
\begin{equation}
\expt{t}{F} = \frac{\gint{F}{-S_t}}{\ginta{-S_t}}
\quad
\text{and}
\quad
\expt{a}{F} = \frac{\gint{F}{-S_a}}{\ginta{-S_a}}\, .
\end{equation}
Before addressing the problem of how to choose an approximate action,
we first need a technique to compare these expectations.
The simplest numerical comparison technique is to generate independent
sequences of configurations for each measure, to evaluate expectations
on these sequences, and then compare.  For our purposes, this simple
approach is not computationally feasible. We are attempting to
find approximate actions
for a given true action because the computational work needed to
generate independent configurations in the true measure is very large.
Our eventual goal is to find a tuning technique which allows us to
generate an approximate action without ever having to simulate the
measure of the true action.  Thus we are lead to consider methods
by which we can use ensembles of configurations generated in one
measure to estimate expectations of operators in a second measure.

Provided the fundamental degrees of freedom of two action functionals,
$S_1$ and $S_2$, are the same (\ie provided the configuration space upon
which they are defined is the same), we can directly relate expectations
in the corresponding measures,
\begin{equation}
\expt{j}{F}
   = \frac{ \expt{i}{ F e^{S_i-S_j} } }{ \expt{i}{ e^{S_i-S_j} } }
   = \expt{i}{F}
     + \expt{i}{ \left(F - \expt{i}{F}\right) C_i(\Delta_{ij}) } \, ,
 \label{Eqn:MeasureComparison}
\end{equation}
where
\begin{equation}
 \Delta_{ij} = S_i - S_j = -\Delta_{ji}
\end{equation}
and
\begin{equation}
C_i(\Delta) =  \frac{1}{\expt{i}{e^\Delta}} e^{\Delta}\, .
\end{equation}
This comparison formula is exact.
However, the calculation
of the  overlap expectations of $C_i(\Delta_{ij})$ with $F-\expt{i}{F}$
is very difficult numerically, since $\Delta_{ij}$ is typically a quantity
which is of $O(V)$, where $V$ is the volume of our system.  If we use
this formula, therefore, to calculate expectations for $S_j$ by correcting
an ensemble of configurations generated for $S_i$, we find that we must
re-weight
with factors which are
$O(e^V)$.  The fluctuations in expectations so calculated are also of
$O(e^V)$, so the work needed to get reasonable statistics is exponential
in $V$.  As a result, a direct application of this formula to compare
$\expt{i}{F}$
and $\expt{j}{F}$ is not feasible numerically.

We adopt instead a cumulant expansion technique which is generated
by expanding $C_i(\Delta)$ in a Taylor series,
\begin{eqnarray}
C_i(\Delta)  & = & 1 + \left( \Delta - \expt{i}{\Delta} \right)
    \nonumber \\
  & & + \frac{1}{2} \left(
         \left( \Delta - \expt{i}{\Delta} \right)^2
         + \expt{i}{\Delta}^2 - \expt{i}{\Delta^2} \right) + \cdots.
\label{Eqn:CumExpans}
\end{eqnarray}
Substituting this expansion in (\ref{Eqn:MeasureComparison}),
we find
\begin{eqnarray}
\expt{j}{F} - \expt{i}{F}
     & = & \expt{i}{ \left(F - \expt{i}{F}\right)
                 \left(\Delta_{ij} - \expt{i}{\Delta_{ij}} \right) }
 \nonumber \\
   & & + \frac{1}{2} \expt{i}{ \left(F - \expt{i}{F}\right)
                             \left(\Delta_{ij} - \expt{i}{\Delta_{ij}} \right)^2 }
     + \cdots.
  \nonumber \\
\label{Eqn:FCumExpans}
\end{eqnarray}
The leading term on the right hand side here is the connected part of
the correlation of the operator $F$ with the difference in the actions
$\Delta_{ij}$.  As before, $\Delta_{ij}$ is a quantity of $O(V)$; but now
we don't have to exponentiate, and so we expect fluctuations only of $O(V)$
rather than $O(e^V)$.  Since any numerical simulation must
be carried out at finite volume, we can expect that at least this leading
correction between the expectation values in the different measures will
be calculable with finite work. Of course, higher order terms will
require calculation of connected correlations involving higher powers
of $\Delta_{ij}$.  These correlations will have correspondingly larger
fluctuations.  The connected correlations involving $\Delta_{ij}^n$ will
have fluctuations of $O(V^n)$, for example, and will be
correspondingly  harder to calculate.  One should view this cumulant expansion
technique, therefore, as the numerical version of an asymptotic expansion.
If the difference $\Delta_{ij}$ is sufficiently small, one or perhaps
two terms in this expansion may be calculable with reasonable work.


\section{Tuning the Approximate Action}
\label{Sec:Tuning}

In our prototypical
example introduced in the last section,
the true action $S_t$ is the lattice QCD action which combines a single
plaquette gauge term and a Wilson fermion determinant term expressed as
the trace of the log of the Wilson hopping matrix.  Our prototypical
approximate action $S_a$ includes just the single plaquette gauge term
but with a possibly different coupling from that in the true action. More
generally, we imagine that the approximate action depends on a finite
number of couplings $a_1, \ldots, a_n$,
\begin{equation}
S_a \equiv S_a(a_1, \ldots, a_n)\, .
\end{equation}
One possibility, for example, is that $S_a$ includes a sum of
different translationally invariant Wilson loop operators \cite{SW1}.
The $a_i$ are then the couplings
multiplying
each different Wilson loop operator in the sum.
Another possibility is to replace the fermion determinant term with
an approximate bosonic term proposed by L\"{u}scher \cite{Luscher}.  The $a_i$
then include the plaquette coupling for the gauge term together with
all the roots of the polynomial used to approximate the inverse of
the fermion matrix in the L\"{u}scher approach.
The question we now address is how to choose the couplings $a_i$ so that
$S_a$ becomes a good approximation to the true action $S_t$.  We define
three different tuning schemes for the coefficients, and show that
these three different schemes are closely related.

\subsection{Minimizing the Distance Norm}
\label{SubSec:DistNorm}
The first tuning procedure is described in \cite{SW1}. There
it is observed
that the measure defined by an action functional $S$ allows us to define an
inner product $(\,,\,)$ on the space of gauge invariant functionals of the link
matrices, which takes the form,
\begin{equation}
(F, G) = \expta{ F^* G } = \frac{\gint{F^* G}{-S}}{\ginta{-S}} \, ,
\end{equation}
where $F$, $G$, and $S$ are all functions of the link matrices
$U_\mu(x)$.  This inner product turns the space of gauge invariant
functions into a Hilbert space, and allows us to define the distance
$d(F,G)$ between any two functions as
\begin{equation}
  d(F,G) = ( (F-G), (F-G) ) = \expta{ (F-G)^*(F-G) }\, .
\end{equation}
Functions of link matrices can now be considered as vectors in this
Hilbert space and the natural tuning condition in this language is
to minimize the distance, or equivalently to maximize the
overlap, between the two vectors which represent the true and approximate
actions.   Since an overall
constant shift in an action functional has no effect on the measure
it generates, the tuning should minimize the connected part only.
Thus, the first prescription is to minimize
\begin{equation}
\label{Eqn:DeltaNorm}
d(S_a - \expta{S_a}, S_t - \expta{S_t})
  = \expta{ (\Delta_{at} - \expta{ \Delta_{at}})^2 }
\end{equation}
over the space of couplings $a_1, \dots, a_n$.
$\Delta_{at}$ is the difference in the approximate
and true actions,
\begin{equation}
  \Delta_{at} = S_a - S_t = - \Delta_{ta}
\label{Eqn:DeltaAT}
\end{equation}
and the object to be minimized is simply the variance
$\sigma^2(\Delta_{at})$ of this difference,
\begin{equation}
\label{Eqn:DeltaVariance}
\sigma^2(\Delta_{at})
  = \expta{ (\Delta_{at} - \expta{ \Delta_{at}})^2 }
\end{equation}
evaluated in the measure $\expta{\ }$.
Differentiating
with respect to each of these couplings in turn, we generate a
set of $n$ simultaneous equations to determine the minimum,
\begin{equation}
 \label{Eqn:MaximizeOverlap}
 \expta{ (\cS_i - \expta{\cS_i})(\Delta_{at} - \expta{\Delta_{at}}) } = 0
  \quad
  \text{ for $i = 1, \dots, n$,}
\end{equation}
where
\begin{equation}
  \cS_i = \frac{\partial}{\partial a_i} S_a(a_1, \dots, a_n)\, .
 \label{Eqn:cSi}
\end{equation}
In deriving this particular tuning condition, we have been
purposefully vague about the measure in which expectations are
to be evaluated.  There are two obvious measures available,
that generated by $S_t$, and that generated by $S_a$.  Either
measure will allow a calculation of the tuned values for the
couplings $a_i$, so this first tuning condition actually has two
possible variants.  Our preferred choice is to minimize in the true
measure.  This minimization is an explicit
problem, but in principle requires generating a set of configurations
in the full measure.  In practice, we could also use the cumulative
expansion methods of the last section to estimate expectations
in the true measure using our best guess as to the
approximate measure.
Minimizing in the approximate measure,
on the other hand, is an implicit problem since we only learn
the correct values for the coefficients $a_i$ after the minimization is
completed.  Minimization in this case requires an iterative
method which first guesses the correct couplings, then evaluates
the corrections to those guesses.

For notational simplicity in what follows, we will assume that
equations of the form (\ref{Eqn:MaximizeOverlap}) apply for
all values of $i$, and we will drop the phrase 
\lq for $i = 1, \dots, n$\rq{} in such equations.

\subsection{Operator Matching}
\label{SubSec:OperatorMatch}
A different approach to tuning the approximate action is suggested by
the observation that, in some cases at least, both true and approximate
actions represent different regularizations of the same continuum
theory.  For the prototypical case of full Wilson fermion action versus
quenched action this is, of course, not correct since the quenched
approximate action does not have a unitary continuum limit.  However, it
is possible to imagine cases where this is indeed true.  For example,
one could consider the case where one action was the full Wilson fermion
action, while the second action was an $O(a)$ improved  action.
In this view, the couplings in both actions are bare parameters in
different regularizations of the same continuum theory.

Normally, one fixes the mapping from bare to renormalized couplings by
specifying renormalization conditions for each bare parameter whose
finite part needs to be fixed.  These conditions directly relate
regularized theories to the renormalized continuum theories which they
regulate.  In our case, imposing renormalization conditions which relate
the regularized theories to continuum renormalized theories is not
feasible, since this requires that we separately generate a full
continuum limit extrapolation of each regularized theory.  What one can
do, however, is to impose renormalization conditions directly between the
two lattice regularized theories.  If these conditions are maintained in
the passage to the continuum limit, then both regularized theories will
proceed to the same renormalized continuum theory as we require.

Consider, therefore, matching the approximate action $S_a(a_1, \ldots,
a_n)$ with the true action $S_t$.  The approximate action has $n$
couplings which must be determined, so we must impose $n$ independent
matching conditions between the approximate and true measures to
fix these $n$ couplings.  The obvious possibilities include matching
operators such as $m_\pi$, $m_\rho$, $m_N$ \cite{Petronzio}.
These
particular operators however are not ideal for our purposes
since they are difficult to measure accurately on a
small number of configurations, and they are not particularly
sensitive to small changes in the couplings $a_i$ which we need to
determine.  The choice of matching
conditions which we make, however, is not really very important since
any
set of
conditions will correctly fix the cutoff dependent divergences of the
couplings being matched, and will differ only in the way they fix finite
parts.  Different matchings will differ by finite renormalizations only,
so we are free to choose some simply implemented matching conditions.
An alternative set which is simple to calculate, and which should be
sensitive to changes in the $a_i$, is to match the operators $\cS_i =
\partial S / \partial a_i$ between the approximate and true actions.

Thus, the second tuning condition we propose is to match
the expectations of the $n$ operators $\cS_i$ ($i = 1, \dots, n$) in
the approximate and true measures,
\begin{equation}
  \expt{t}{\cS_i} = \expt{a}{\cS_i}.
\end{equation}
Applying the cumulant expansion to $\expt{a}{S_i}$, we find
\begin{equation}
  \expt{t}{\cS_i} = \expt{t}{\cS_i}
                  + \expt{t}{(\cS_i-\expt{t}{\cS_i})
                               C_t(-\Delta_{at})} \, ,
\end{equation}
which directly gives
\begin{equation}
  \expt{t}{(\cS_i-\expt{t}{\cS_i}) C_t(-\Delta_{at})}
                  = 0\, .
  \label{Eqn:MatchCondition}
\end{equation}

\subsection{Maximizing Acceptance}
\label{SubSec:MaximizeAccept}

Consider using an approximate action as part of an exact Markov process
which generates configurations distributed according to the measure of
the true action.  The particular exact algorithm we propose
involves a standard Markov transition step which satisfies detailed
balance for the approximate action.  This step generates new trial
configurations which do not have the correct true action distribution.
One can then correct to the true distribution by executing a Metropolis
accept/reject on the trial configurations.  Our third tuning condition
for the approximate action is to maximize acceptance in this exact
algorithm.

Let $P_a(U \rightarrow U')$ be a Markov transition probability for
link configuration $U$ to go (reversibly) to link configuration $U'$.  If this
transition probability satisfies detailed balance for the approximate
action $S_a$, then
\begin{equation}
\left. e^{-S_a} \right|_{U} P_a(U \rightarrow U')
 = \left. e^{-S_a} \right|_{U'} P_a(U' \rightarrow U) \, ,
\end{equation}
where the notation $\left.\right|_{U,U'}$ indicates that the relevant
action terms are evaluated on the link configurations labeled by $U$
and $U'$.  To get an exact algorithm for the true action $S_t$ we must
now execute an accept/reject step. The
probability, $A_{at}$, to accept a trial step generated by $P_a$ must
satisfy a modified detailed balance,
\begin{equation}
\left. e^{\Delta_{at}} \right|_{U} A_{at}(U \rightarrow U')
 = \left. e^{\Delta_{at}} \right|_{U'} A_{at}(U' \rightarrow U)
\end{equation}
and the optimum choice for this acceptance probability is
\begin{equation}
A_{at}(U\rightarrow U') = \min\left( 1, e^{\delta_{U\rightarrow U'}(\Delta_{at})} \right)
\end{equation}
where
\begin{equation}
\delta_{U\rightarrow U'}(\Delta_{at})  = \left. \Delta_{at} \right|_{U'}
                       - \left. \Delta_{at} \right|_{U}\, .
\end{equation}
Maximizing acceptance for this compound algorithm requires maximizing
the weighted average of $A_{at}$ over all configurations $U$ which can
occur at the start of a trial step, and all configurations $U'$ which
can occur at the end of such a step.
If we denote this weighted average as $\dexpt{ A_{at} }$, we have
\begin{equation}
\dexpt{ A_{at} } = \int d[U'] d[U]
              \; A_{at} \;
              \left. \frac{e^{-S_t}}{\int d[U] e^{-S_t}} \right|_U
              P_a(U \rightarrow U').
\end{equation}
Since $P_a$ satisfies detailed balance for action $S_a$, it is
straightforward to show that
\begin{equation}
e^{\dexpt{\delta_{U\rightarrow U'}(\Delta_{at})}}
 \quad \le \quad \dexpt{ e^{\delta_{U\rightarrow U'}(\Delta_{at})} }  =  1
\end{equation}
which implies
\begin{equation}
\dexpt{\delta_{U\rightarrow U'}(\Delta_{at})} \quad \le \quad 0.
\end{equation}
Further, if $\dexpt{ \delta(\Delta_{at})} $ is small, and if only the
leading terms in the Taylor expansion of
$e^{\dexpt{\delta(\Delta_{at})}}$
are significant, the acceptance probability is given approximately by \cite{deForc}
\begin{equation}
\label{Eqn:Erfc}
\dexpt{ A_{at} } \approx \hbox{erfc} \left(\frac{1}{2}
                              \sqrt{ - \dexpt{\delta(\Delta_{at}) }
	} \right)\, .
\end{equation}
Acceptance, therefore, is maximized when $-\dexpt{\delta(\Delta_{at})}$
is minimized.
The worst case for acceptance occurs when the transition
probability $P_a(U\rightarrow U')$ defines a perfect heat bath
transition for the approximate action,
\begin{equation}
P_a(U\rightarrow U') \rightarrow
  \left. \frac{e^{-S_a}}{\int d[U] e^{-S_a}} \right|_{U'}.
\end{equation}
Trial configurations $U'$ in this case are completely independent of
the configurations $U$ which they attempt to replace.
In this worst case, we have:
\begin{eqnarray}
\label{Eqn:DeltaDeltaExpt}
-\dexpt{\delta(\Delta_{at})} & = &
  -\dexpt{ \left. \Delta_{at} \right|_{U'} - \left. \Delta_{at} \right|_U }
  \nonumber\\
  & = & - \left( \expt{a}{\Delta_{at}} - \expt{t}{\Delta_{at}} \right)
  \nonumber\\
  & = & - \expt{t}{ (\Delta_{at} - \expt{t}{\Delta_{at}}) C_t(-\Delta_{at})
	}\, .
\end{eqnarray}
Maximizing acceptance requires maximizing this last term over the
couplings $a_1, \dots, a_n$ in the approximate action $S_a$.
If we differentiate with respect to the couplings $a_i$,
we find the following condition:
\begin{equation}
\expt{t}{(\cS_i-\expt{t}{\cS_i}) C_t(-\Delta_{at}) }
- \expt{t}{(\cS_i-\expt{t}{\cS_i C_t(-\Delta_{at})})
                  (\Delta_{at} - \expt{t}{\Delta_{at}})
                  C_t(-\Delta_{at}) }
                   =  0.
  \label{Eqn:MaximizeAccept}
\end{equation}

An important question to ask once an approximate action has been tuned
to maximize acceptance is whether the acceptance achieved by this
maximization is large enough to generate a practical exact algorithm.
The answer, of course, is given by the value of $\dexpt{A_{at}}$ which
results.  $\dexpt{A_{at}}$ is given in (\ref{Eqn:Erfc}).  We can
rearrange this equation by using the worst case formula for
$-\dexpt{\delta(\Delta_{at})}$ given in
(\ref{Eqn:DeltaDeltaExpt}).  When expanded, this latter equation takes the very
simple form,
\begin{eqnarray}
\label{Eqn:deltadeltaexpanded}
-\dexpt{\delta(\Delta_{at})}
& = &
 \expt{t}{ (\Delta_{at} - \expt{t}{\Delta_{at}})^2 }
 + O(\Delta_{at}^3)
 \nonumber \\
& = & \sigma^2_t(\Delta_{at}) + O(\Delta_{at}^3) \, .
\end{eqnarray}
(The subscript $t$ on $\sigma_t^2$ denotes that the expectations
involved are to be evaluated in the true action measure).
Thus, we find
\begin{equation}
\label{Eqn:Erfc2}
\dexpt{ A_{at} } \approx
\hbox{erfc} \left(\frac{1}{2}\sqrt{ - \dexpt{\delta(\Delta_{at}) }} \right)
\approx
\hbox{erfc} \left(\frac{1}{2}\sigma_t(\Delta_{at})\right)\, .
\end{equation}

The actual minimum value for $\dexpt{A_{at}}$ which defines a
practical algorithm is a function of the ratio of work involved in
generating uncorrelated trial configurations with approximate and true
actions.  In QCD, the generation of uncorrelated configurations with,
for example, an
exact hybrid algorithm is very expensive.  This suggests that even very
small acceptances might suffice to generate practical exact algorithms
based on approximate actions.  A 50\% acceptance is achieved when
$\sigma^2_t(\Delta_{at}) \approx 1$, while a 10\% acceptance is achieved when
$\sigma^2_t(\Delta_{at}) \approx 5$.

\subsection{Tuning Procedure}
\label{SubSec:TuningProcedure}

We have, at this point, proposed three different tuning conditions,
(\ref{Eqn:MaximizeOverlap}, \ref{Eqn:MatchCondition}, \ref{Eqn:MaximizeAccept}),
to define the adjustable parameters $a_i$ in an approximate
action.
Two of these equations involve the cumulant
expansion factor $C_t(\pm\Delta_{at})$.  As we have argued in the last
section, this factor is not directly calculable in a single numerical
experiment with a single ensemble generated by either the
approximate or true action.  Instead, to
apply these tuning conditions, we
must expand $C_t(\pm\Delta_{at})$ in powers of $\Delta_{at}$
using (\ref{Eqn:CumExpans}).
The tuning conditions which result when we keep just the leading
and next to leading terms of this expansion take the following forms.
First, if vector overlap in the true measure is maximized using
(\ref{Eqn:MaximizeOverlap}), we have unchanged
\widetext
\begin{equation}
 \expt{t}{ (\cS_i - \expt{t}{\cS_i})(\Delta_{at} -
 \expt{t}{\Delta_{at}}) }
 = 0 \, .
\end{equation}
Second, if operators are matched in approximate and true measures
(\ref{Eqn:MatchCondition}),
then we have
\begin{equation}
  \expt{t}{(\cS_i-\expt{t}{\cS_i})(\Delta_{at}-\expt{t}{\Delta_{at}})}
  -\frac{1}{2}\expt{t}{(\cS_i-\expt{t}{\cS_i})
                       (\Delta_{at}-\expt{t}{\Delta_{at}})^2 }
  + \cdots =  0 \, .
\end{equation}
Finally, if acceptance is maximized using (\ref{Eqn:MaximizeAccept}), we have
\begin{equation}
 \expt{t}{ (\cS_i - \expt{t}{\cS_i})(\Delta_{at} - \expt{t}{\Delta_{at}}) }
  - \frac{3}{4}\expt{t}{ (\cS_i - \expt{t}{\cS_i})(\Delta_{at} - \expt{t}{\Delta_{at}})^2 }
 +  \cdots = 0 \, .
\end{equation}
\narrowtext
These three conditions all agree to leading order in $\Delta_{at}$,
and only begin to disagree at next to leading order in the cumulant
expansion.  If we have managed to find approximate and true actions
which agree sufficiently closely that the cumulant expansion
is a numerically useful tool for comparing expectations
in the two measures, then we would expect the terms
at next to leading order in these equations to be much smaller
than the leading order terms.

These observations suggests the following self-consistent approach to generating
approximate actions for lattice simulation.
First, apply one of the three conditions listed here to tune
the effective action.  This tuning can be carried out for an ensemble
of configurations generated either for the approximate or for the full
measure, so long as the cumulant expansion is used appropriately.
Once the approximate action is determined, a variety of operators should
be calculated in this approximate action, and further, the shift between
expectations in approximate and true actions should be determined for
these operators using the cumulant expansion.
If the shift in operators so calculated is small, then we expect that
the approximate action provides a good representation of the true
action.  Further, we expect that all three possible tuning conditions
will be approximately satisfied, and that the cumulant expansion will
work well for most operators.
On the other hand, if the shift in the operators so calculated is large, then
the approximate action is not a good representation of the true action.
In this case, we find that the different matching conditions will
produce significantly different values for the approximate action
couplings.  The situation which will then prevail is that we will be
able to adjust certain operators to match between the approximate and
true measures.  However, this adjustment will not cause other different
operators to match, and simulation with the approximate action will not provide
a useful approximate ensemble for the study of properties of the true action
ensemble.

At leading order in the cumulant expansion, the common tuning
condition from all three methods takes the form
\begin{equation}
 \expt{t}{ (\cS_i - \expt{t}{\cS_i})(\Delta_{at} -
 \expt{t}{\Delta_{at}}) } = 0\, .
\end{equation}
As written, this condition requires that expectations are to be
evaluated in the true measure.  As we suggested above, we can
use the cumulant expansion to apply this condition using operators
evaluated in the approximate measure also, so long as we use the
cumulant expansion to correct appropriately.
We have
\begin{eqnarray}
\label{Eqn:LeadingTuning}
 \lefteqn{\expt{t}{(\cS_i-\expt{t}{\cS_i})
                   (\Delta_{at}-\expt{t}{\Delta_{at}})} \quad}
 \nonumber \\ & = & \expt{a}{(\cS_i-\expt{t}{\cS_i})
              (\Delta_{at}-\expt{a}{\Delta_{at}})}
 +
 \expt{a}{O(\Delta_{at}^2)}.  
\end{eqnarray}
These two forms differ only by terms which we have already
neglected when we work to leading order.  Thus we can
quite consistently
choose to drop the next to leading terms on the right hand side
of (\ref{Eqn:LeadingTuning}) when we wish to
use an approximate action ensemble rather than a true action
ensemble to perform tuning.

Our leading order approximate action tuning
condition now takes the simple form
\begin{equation}
 \label{Eqn:MinCondition}
 \expta{ (\cS_i - \expta{\cS_i})(\Delta_{at} -
 \expta{\Delta_{at}}) } = 0
\end{equation}
where expectations are evaluated in either true or approximate measures.
This condition represents a set of coupled equations for the
tunable coefficients $a_i$ in the approximate action $S_a$,
and are precisely
the equations which minimize
$\sigma^2(\Delta_{at})$
over these
coefficients.  $\sigma^2(\Delta_{at})$ therefore
provides a key measure of the quality of an approximate action.
It also, according to (\ref{Eqn:deltadeltaexpanded}), determines whether a practical
exact algorithm can be built using a global accept/reject step to correct
from approximate to true action.
In what follows, we use this variance to discriminate between candidate
approximate actions and to guide our search for an exact update
algorithm based on the approximate action.

\subsection{Wilson Loop Approximations}
\label{Sec:WilsonLoops}

To close this section, let us consider the technicalities of
minimizing $\sigma^2$.
The approximate actions which we
study in the remainder of this paper are simple sums of
Wilson loop operators of different sizes and shapes.
We will use the notation $\cW_i$ to denote a particular Wilson loop
operator which we always take to be translationally and rotationally
invariant.  A typical example of such an operator in QCD is the
$1\times 1$ plaquette summed over all sites and orientations,
\begin{equation}
\cW_{1\times 1} = \sum_\Box \frac{1}{3} \Real \Tr U_\Box\, .
\end{equation}
Wilson loop operators are always summed  over all distinct orientations
and locations on the lattice.  Thus the operators we consider are
rotationally and translationally invariant.  This is sufficient for our
purposes since our major effort is to find an approximation for the
fermion
determinant which is also rotationally and translationally invariant.

The class of approximate actions which we will consider now takes
the simple form
\begin{equation}
S_a(a_1, \dots, a_n) = -\sum_{i=1}^n a_i \cW_i,
\end{equation}
and our goal will be to find the best approximation within this
class to the true action
\begin{equation}
S_t = -\beta \cW_{1\times 1}  - T
\end{equation}
where $T$, defined as,
\begin{equation}
T = \frac{n_f}{2} \Tr \log M^\dagger M
\end{equation}
is the trace of the log of the fermion determinant for $n_f$ equal
mass fermions.
We have immediately that
\begin{equation}
\frac{\partial}{\partial a_i} S_a(a_1, \dots, a_n)
 =\cS_i =-\cW_i\, .
\end{equation}

The difference between approximate and true actions for this case
is
\begin{eqnarray}
\Delta_{at} & = & S_a - S_t \nonumber \\
            & = & T + \beta \cW_{1\times 1} - \sum_i a_i \cW_i \nonumber \\
            & = & T - \sum_1 a'_i \cW_i
\end{eqnarray}
where, for convenience, we have absorbed the pure gauge term by
redefining the coefficients $a'_i$,
\begin{eqnarray}
     a_{1\times 1}' & = & a_{1\times 1} - \beta  \\
     a_i' & = & a_i \qquad\qquad i \ne 1\times 1.
\end{eqnarray}
and the work of finding an approximate action reduces to that of finding
an approximation for $T$ of the form $T \approx \sum_i a_i \cW_i$.
For convenience in what follows,
we drop the prime superscript on the parameters $a_i$
which define this approximation.
Our lowest order minimization condition (\ref{Eqn:MinCondition}) is now
simply expressed as
\begin{equation}
\label{Eqn:LinearSystem}
\sum_{j=1}^n \expta{ (\cW_i - \expta{\cW_i}) (\cW_j - \expta{\cW_j}) } a_j
 = \expta{(\cW_i - \expta{\cW_i})(T -  \expta{T})}\, .
\end{equation}
To determine the parameters $a_i$ which define the best approximation
to $T$ we must calculate connected correlations of each of the
Wilson loop operators included in the approximate action both
with themselves
(\ie $\expta{ (\cW_i\!-\!\expta{\cW_i})(\cW_j\!-\!\expta{\cW_j})}$),
and with $T$
(\ie $\expta{ (\cW_i\!-\!\expta{\cW_i})(T\!-\!\expta{T})}$).
Once these correlators are available, then the evaluation
of the $a_i$'s is achieved by
solving the linear system (\ref{Eqn:LinearSystem}).

It is also useful, in the process of generating the coefficients
$a_i$ of the approximation to $T$, to define a Gram-Schmidt
orthogonalized basis of Wilson loop operators.
If the original loops are $\cW_i$ for $i = 1, \dots, n$, then
the orthogonalized loops are $\hat{\cW_i}$ and are defined
recursively by
\begin{equation}
\hat\cW_1=\cW_1-<\cW_1>
\label{Eqn:S1hat}
\end{equation}
and
\begin{equation}
\hat\cW_i=\cW_i-<\cW_i>-\sum_{j=1}^{i-1}{C_i}^j\hat\cW_j\,,
\qquad\, i>1\, ,
\end{equation}
where
\begin{equation}
{C_i}^j= \frac{\expta{\cW_i\hat\cW_j}}{\expta{\hat\cW_j^2}}\, .
\end{equation}
The operators $\hat\cW_i$ therefore satisfy
\begin{equation}
\expta{\hat\cW_i\hat\cW_j}
=\expta{\hat\cW^2_j}\delta_{ij}\, ,\quad \expta{\hat\cW_i}=0\,.
\label{Eqn:Orthog}
\end{equation}
and, in terms of these orthogonalized operators, the solution of the
linear system (\ref{Eqn:LinearSystem}) is given by
\begin{equation}
a_i={<{\hat\cW_i(T-<T>)>}\over{<\hat\cW_i^2}>}\, .
\label{Eqn:a_i}
\end{equation}


\section{The Schwinger Model}
\label{Sec:SchwingerModel}

In order to test the ideas developed in the proceeding section, we need
a model which exhibits the important features of the QCD lattice model,
but is computationally accessible.  The
Schwinger model offers a
suitable testing ground.  This theory is a two-dimensional gauge theory
with gauge group $U(1)$ rather than $SU(3)$.  With no fermions present
the theory is (almost) trivial.  All the interesting dynamics are
generated by fermions, and we expect very significant dynamical fermion
effects to be present as a result.  Computationally, it is possible to
generate full dynamical fermion simulations in just a few days on a
workstation, so the model allows considerable experimentation with
multiple different actions over reasonable time scales.  The fact that
much is known exactly and perturbatively about the corresponding
continuum theory provides an important bonus.

The explicit form of Schwinger Model action which we adopt is
\begin{equation}
\label{Eqn:SchwAction}
S = - \beta \sum_\Box  \Real U_\Box
  - \frac{n_f}{2} \Tr \log M_\kappa^\dagger M_\kappa.
\end{equation}
$n_f$ is the number of equal mass fermions included, which we
take to be 2 in all our simulations.
Gauge elements $U_\mu(x)$ are defined on the links of a two dimensional
square lattice and take values in the group $U(1)$.
The hopping matrix $M_\kappa$ is given by the usual Wilson form
\begin{eqnarray}
M_\kappa(x,y) & = & \delta_{x,y} - \kappa K(x,y),
\\
K(x,y) & = &
  \left( (1-\gamma_\mu) U_\mu(x) \delta_{x+\hat{\mu},y}
       + (1+\gamma_\mu) U^\dagger_\mu(x-\hat{\mu}) \delta_{x-\hat{\mu},y}
       \right).
\end{eqnarray}
$M$ can be decomposed
into its red/black (even/odd) components,
\begin{equation}
\label{Eqn:FermionMatrix}
M \equiv 1 - \kappa K
= \left( \begin{array}{cc}
         1 & -\kappa  K^{oe} \\
         -\kappa K^{eo} & 1 \end{array}
         \right)\, .
\end{equation}
Using this decomposition, one may construct a Hermitian, positive definite,
red/black preconditioned matrix $H^{(e)}$ such that
\begin{equation}
H^{(e)}=M^{(e)}M^{(e)\dagger}\, , \qquad
M^{(e)}=1-\kappa^2K^{eo}K^{oe} \,.
\label{Eqn:RedBlack}
\end{equation}
For two flavours of equal mass fermions,
the fermion contribution to the Schwinger model action is then given
by
\begin{equation}
\label{Eqn:TDef}
T = \Tr \log M^{\dagger}_\kappa M_\kappa = \Tr \log H^{(e)}\, .
\end{equation}
In what is to follow, we will use preconditioned iterative methods to
calculate inverses and logs of the fermion matrix.  We have found that
the red-black form here provides the best preconditioning.

In the language of the last section, our true action is the
Schwinger model action with dynamical fermions included and is given by
(\ref{Eqn:SchwAction}).  We used the hybrid molecular dynamics algorithm
\cite{HMC} with a five step integration scheme \cite{SW3}
to generate ensembles of configurations distributed according to
this true action.
All the approximate actions which we considered
involved varying sets of Wilson loop operators.
To generate ensembles distributed according to these
actions, we used a simple Metropolis link update algorithm.

The only other simulation issue worth commenting upon is that of the
calculation of $T$ and $T^2$ for individual configurations in an
ensemble.  These calculations are required whenever we need to find
correlations between action terms.
$T$ is
defined in (\ref{Eqn:TDef}) as the trace of the log of the matrix
$H^{(e)}$.  To calculate $T$ we adopt a stochastic approach
\cite{SW1}.  A number $N_\phi$ of Gaussian distributed vectors $\phi_a(x)$
are generated.  The index $x$ here runs over the (even) sites of the
lattice, and $a = 1, \dots, N_\phi$ labels the different vectors
generated.  We use the notation $\expt{\phi}{\ }$ to denote
expectations over the vectors $\phi$ and the
probability distribution of the components of these
vectors are normalized so
\begin{equation}
\expt{\phi}{ \phi_a(x) \phi^\dagger_b(y) } = \delta_{a,b} \delta_{x,y}\, .
\end{equation}
The stochastic estimators, $E_T$ and $E_{T^2}$, which we adopted for
$T$ and $T^2$ respectively, are given by
\begin{eqnarray}
E_T & = & \frac{1}{N_\phi} \sum_a \left(\phi^\dagger_a L \phi_a\right)
\\
E_{T^2} & = & \frac{1}{N_\phi (N_\phi - 1)}
\sum_{a\ne b} \left(\phi^\dagger_a L\phi_a\right)  \left(\phi^\dagger_b L \phi_b\right)
\end{eqnarray}
where, for convenience, we define
\begin{equation}
L  =  \log H^{(e)}\, .
\end{equation}
We have immediately that
\begin{eqnarray}
\expt{\phi}{E_T} & = & \Tr L = T \\
\expt{\phi}{E_{T^2}} & = & (\Tr L)^2 = T^2 \\
\expt{\phi}{(E_T)^2} & = & (\Tr L)^2 + \frac{2}{N_\phi} \Tr (L^2) \, .
\end{eqnarray}
Note that $E_{T^2}$ as defined here is an unbiased estimator of $T^2$.
The other obvious choice of estimator for $T^2$ is $(E_T)^2$.
We see from these equations that $E_T^2$ is a biased estimator for $T^2$ which
only becomes unbiased in the limit $N_\phi\rightarrow\infty$.
The variances of the estimators $E_T$ and $E_{T^2}$ are also easy
to calculate.  We have
\begin{equation}
\expt{\phi}{(E_t- \expt{\phi}{E_T})^2}
= \frac{2}{N_\phi} \Tr (L^2)\, .
\end{equation}
The variance in $E_{T^2}$ has a more complicated form, but a similar
$1/N_\phi$ overall factor.  When using these estimators to calculate $T$
and $T^2$, we must therefore choose $N_\phi$ sufficiently large that the
fluctuations due to $\phi$ are small.

The remaining technical problem is to calculate
$L \phi = \log H^{(e)} \phi$.  We use a Chebychev polynomial
approximation (of order $N_C$)
to perform this
calculation.  To optimize the polynomial approximation used, it is
necessary
to know the maximum  and minimum eigenvalues of the matrix $H^{(e)}$.
These
values are obtained with a Lanczos method \cite{Golub}.  The calculation
of the Chebychev polynomial approximation itself then requires
$N_C$  matrix-times-vector products for each vector $\phi$.
In all our analyses, we have found that
the time needed to estimate $T$ and $T^2$ is a significant fraction (5\%-50\%)
of the time needed to perform a full dynamical simulation.

Arbitrary numbers of equal mass fermions can, of course, be dealt with
by incorporating a factor $n_f/2$ along with $T$ in the accompanying
formalism.


\section{Lowest Order Results}
\label{Sec:SchwingerResults}

\subsection{The Quenched Approximation}

Our preliminary application of the methods introduced in
Section \ref{Sec:Tuning} to the Schwinger model
was performed on lattices of size $16^2$ and $32^2$, at
$\beta$ values from 2.0 to 3.0, and $\kappa$  from $0.25$ to $0.265$.
Typically, we generated 1000 or 2000 configurations using
either the true full fermion action or an approximate action.
Maximum and minimum eigenvalues of $H^{(e)}$ were calculated
on each of these configurations using a Lanczos algorithm.
The major computational effort is then to evaluate the stochastic estimators
$E_T$ and $E_{T^2}$ of $T$ and $T^2$ respectively on each configuration.
We used 30-50 different Gaussian $\phi$'s
per configuration to accumulate these estimates, and
we used Chebychev polynomials of order 100-200
to calculate $L \phi \equiv \log H^{(e)} \phi$ for each $\phi$.
We normally worked with configurations which were first fixed to Landau
gauge. This was expected to reduce the influence of gauge artifacts in
the Lanczos procedure and allowed the use of Fourier acceleration
methods. In practice, we found that preconditioning additional to that
provided by the red/black construction gave little extra benefit.

We also accumulated measurements of a variety
of different Wilson loop operators on each  configuration.
As described in Section \ref{SubSec:TuningProcedure},
we will use the notation $\cW_i$ for such a loop operator
labeled by the generic index $i$.
Where appropriate, the generic index
will be replaced by a descriptive index.  For example, $\cW_{1\times 1}$
denotes the simple $1\times 1$ plaquette Wilson loop operator which,
for the Schwinger model, is defined by
\begin{equation}
\cW_{1\times 1} = \sum_\Box \Real U_\Box\, .
\end{equation}
Typical loop operators considered include $1\times 1$,
$2\times 1$, $2\times 2$, $3\times 1$ etc.

When the raw data for the estimators $E_T$, $E_{T^2}$ and the loop operators
$\cW_i$ have been calculated on a configuration by
configuration basis, the final analysis required is to evaluate correlations
of different loop operators with $E_T$, and to derive the values
of the approximate action coefficients according to the conditions
of Section \ref{SubSec:TuningProcedure}.

The first and simplest case to consider is the quenched approximation.
The true action (for two flavours of equal mass fermions,
(\ref{Eqn:SchwAction}))
is 
\begin{equation}
\label{Eqn:SchTrue}
   S_t = - \beta \sum_{\Box} \Real U_\Box - \Tr\log M^\dagger M
       = - \beta \cW_{1\times 1} - T\, .
\end{equation}
The approximate action which defines the quenched approximation
includes just the single plaquette Wilson loop operator, 
$\cW_{1\times 1}$.
The parameters of this approximate action
must however be adjusted according to the
conditions in Section \ref{SubSec:TuningProcedure} and so
we write it with
a different coupling, $\beta'$.
Thus,
\begin{equation}
\label{Eqn:SchApprox1}
S_a =  - \beta' \cW_{1\times1}
\end{equation}
The difference between these two actions is 
\begin{equation}
\Delta \equiv \Delta_{at} = S_a - S_t = T + \delta\beta \cW_{1\times 1}
\end{equation}
where
\begin{equation}
\delta\beta = \beta - \beta'\, .
\label{Eqn:DBeta}
\end{equation}
$\delta\beta$ is the shift in plaquette coupling induced by
approximating
$T$ with a plaquette term only.
Our lowest order tuning condition
is that the variance, $\sigma^2(\Delta)$,
is to be minimized in either the true or approximate gauge measure.
Explicitly we have
\begin{equation}
\sigma^2(\Delta)
= \expta{ \left( (T-\expta{T})
                + \delta\beta
                 (\cW_{1\times 1} - \expta{\cW_{1\times 1}}) \right)^2
		}\, .
\end{equation}
The condition on $\delta\beta$ to minimize this expectation is
\begin{equation}
\delta\beta = -\frac
{\expta{(\cW_{1\times 1} - \expta{\cW_{1\times 1}})(T - \expta{T})}}
{\expta{(\cW_{1\times 1} - \expta{\cW_{1\times 1}})^2}}\, .
\end{equation}
In the initial test, we used the quenched algorithm to generate
configurations.  Thus the ensemble which we have generated is
the appropriate one for the approximate action, and we choose to
minimize the variance of $\Delta$ in this measure.
For $\beta'=2.50$ and $\kappa=0.26$,
we found $\delta\beta=-0.212\pm 0.010$.
These and all other errors were
evaluated  using the bootstrap method,
with subensembling to detect and remove
autocorrelations.
The interpretation of this result is that,
within the space of quenched actions defined by 
(\ref{Eqn:SchApprox1}), the best approximation to
the true action (\ref{Eqn:SchTrue}) with
$\beta = 2.500 - 0.212 (\pm 0.010)$ and $\kappa = .26$ is 
given
by the quenched action
with $\beta'=2.500$.

Having defined an approximate action, 
it is immediately important to check the quality of the
approximation which that action generates. 
In particular, we ask whether we have been justified 
in using just the leading term in the cumulant
expansion when tuning the approximate action.  To investigate this, we
generated a new ensemble of 2000 configurations
using the hybrid molecular dynamics algorithm to incorporate
dynamical fermions ($\kappa=0.26$)
at $\beta=\beta'+\delta\beta=2.29$.
As a simple preliminary test, we compared $\expta{\cW_{1\times 1}}$
as measured using the approximate and true actions.
The quenched, approximate action value was
\begin{equation}
\expt{a}{\cW_{1\times 1}} = 783.3\pm 0.3   \qquad \beta' = 2.50
\end{equation}
while the true dynamical fermion action value was
\begin{equation}
\expt{t}{\cW_{1\times 1}} = 782.8\pm 0.3  \pm 1.4 \qquad \beta = 2.29,
\kappa = 0.26\, .
\end{equation}
The error $\pm 1.4$ in this last equation is due to the uncertainty with
which we have determined $\delta\beta$.
The agreement between the two measurements
is a very encouraging result.  The expectations
of $\cW_{1\times 1}$ match in the true and approximate measures.
We are therefore in the situation where tuning using the
common lowest order tuning conditions (\ref{Eqn:MinCondition})
produces an approximate action which matches the expectation
of an operator in the approximate and true measures.  This is
our first evidence that, in this instance,  we have been justified in taking only
the leading term in the cumulant expansion.

\subsection{Bound state mass}

We now investigate how well the quenched approximation
performs in calculating expectations of
some interesting physical operators in the Schwinger model.
The continuum model with a single species of
massless fermions has a \lq vector\rq{}
bound state with mass
$e/\sqrt{\pi}$ where $e$ is the fermion charge.
For $n_f$ flavours of massless \lq quarks\rq{}, there
is an $SU(n_f)$ symmetry, leading to
an iso-singlet \lq vector\rq{} particle
with mass
\begin{equation}
m_v=e\sqrt{{n_f}\over{\pi}}
\end{equation}
and a massless
$(n_f^2-1)$-plet \cite{Halpern}.
On the lattice we have two flavours and look for a bound state
by studying the large Euclidean time dependence of zero momentum
correlators involving $\overline{\psi}\gamma_1\psi$.
The latter were constructed from \lq quark\rq{} propagators which were
obtained by a Conjugate Gradient solver
applied to the red/black preconditioned Hermitian matrix $H^{(e)}$
or by the BiCGStab algorithm \cite{BiCGStab} used with $M^{(e)}$.
The latter turned out to be the more efficient method.

\begin{figure}[t]
\medskip
\centerline{\psfig{figure=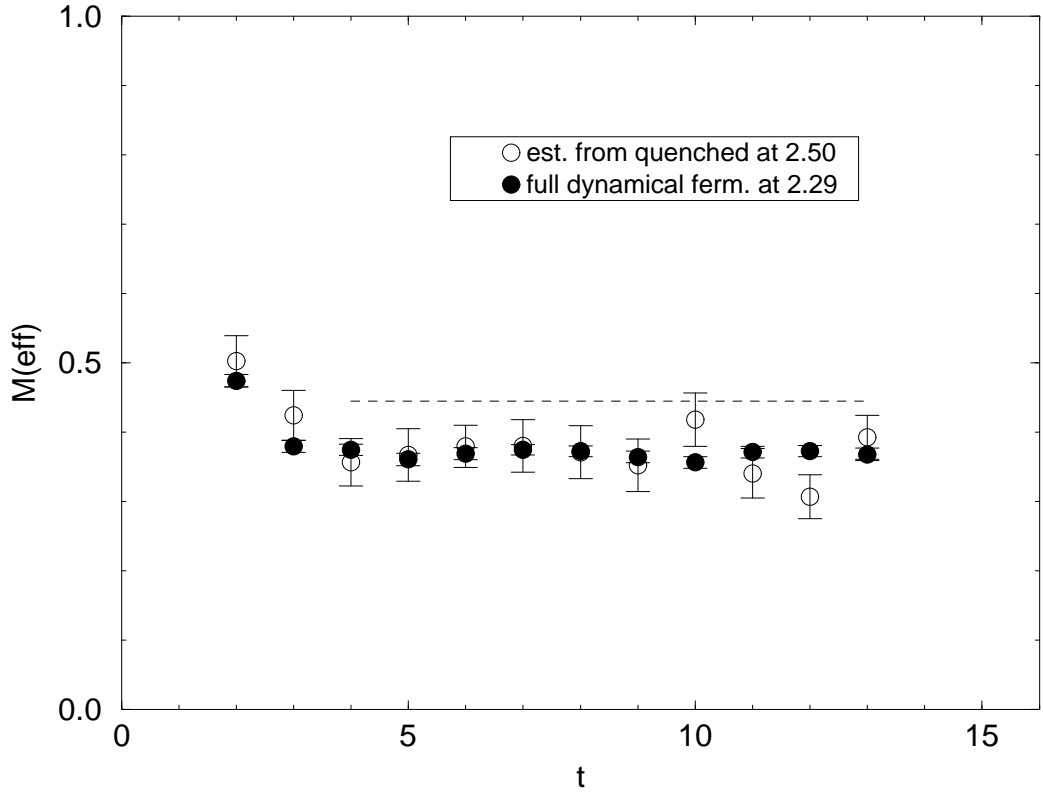,width=5.0in,angle=-90}}
\medskip
\caption{Effective meson masses in the Schwinger model calculated
with a true action (full dynamical fermion) simulation 
($\beta=2.29, \kappa=0.26$) and with a
quenched approximate action simulation at $\beta= 2.50$.}
\label{Fig:EFM}
\end{figure}

Figure \ref{Fig:EFM} shows the vector state effective mass $m_v$
obtained from a full dynamical fermion simulation with the true action at
$\beta=2.29$, $\kappa=0.26$
compared with that obtained from a quenched simulation at
$\beta'=2.50$.
The quenched values plotted for $m_v$ include the leading
\lq deficit  correction\rq{} which is defined by the first 
term on the right hand side of the cumulant expansion
formula (\ref{Eqn:FCumExpans}).
The dashed line in this figure serves both to indicate
the (uncorrected) value for $m_v$ obtained by a quenched
simulation with  $\beta'=2.29$,  and the
Euclidean time range used to extract mass estimates.
From the true action simulation and
from the deficit corrected estimate from the quenched simulation respectively, we find
$m_v=0.369(3)$ and $m_v=0.367(7)$ where the bracketed digit
is the error in the last significant digit.
These vales are in complete agreement.
In the deficit corrected value,
the contribution due to the $\beta'$ shift from $\beta'=2.29$ to
$\beta'=2.5$ is $-.041$ and that 
due to the deficit correction term is $-.037$.

This is pleasing but not yet surprising. The approximate action
simulation
is again generating predictions which are in agreement with those from
the correct true action simulation.  It must, however, be pointed out
that the statistical fluctuations in the approximate action
estimates are, for the same
number of configurations, somewhat larger ($\times 2.5$). So any
comparison of algorithm efficiency must take this into account.  We
will return to this issue later.

The relatively small difference between quenched and full theory estimates
in this particular observable is in accord with experience gained in lattice
studies of the hadron spectrum. It also means that there is little point
in using $m_v$ when studying more complicated approximate
actions. To really stretch the method we now study an observable which is explicitly
dependent on the long distance dynamics.

\subsection{Static potential}

\begin{figure}[t]
\medskip
\centerline{\psfig{figure=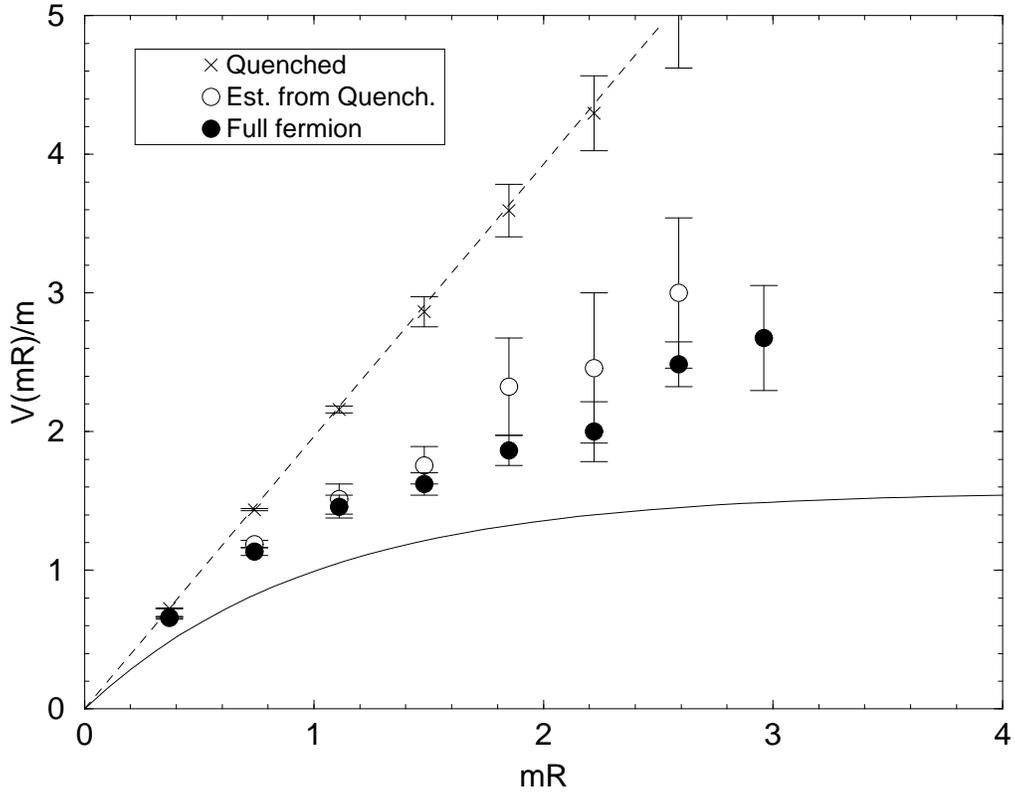,width=5.0in,angle=-90}}
\medskip
\caption{
The static potential in physical units calculated in three different
ways. Crosses indicate the quenched, $\beta=2.50$ result,
open circles indicate the deficit corrected quenched results and
the filled circles are the full dynamical fermion results for $\beta=2.29, \kappa=0.25$.
The curves are described in the text.}
\label{Fig:PhysPot}
\end{figure}

Figure \ref{Fig:PhysPot} shows a comparison
of the static potential  using quenched simulation data at $\beta'=2.50$, full
dynamical simulation data
at $\beta=2.29$, $\kappa=0.26$ and the deficit corrected estimate from the
quenched data. The
results have been expressed in dimensionless units using our result
$m_v=0.37$ in lattice units.
Also shown is the (exact) infinite
volume quenched result (dashed line) and the continuum  massless
Schwinger result (full line)  obtained by evaluating Wilson loops due to
a vector field of mass $m_v$. The continuum result is
\begin{equation}
{{V(m_vR)}\over{m_v}}={\pi\over 2}(1-e^{-m_vR})\, .
\label{Eqn:ContV}
\end{equation}
Note that, expressed this way, the continuum result is independent
of the number of flavours. The comparison between the lattice estimates
is satisfactory although the errors associated with the deficit estimate
(\ref{Eqn:FCumExpans}) get quite large at large distances. This could
signal
a
systematic under-estimate of fermion effects at increasing $R$,
but it is not yet
statistically significant.
The same behaviour is seen in a direct
comparison using Wilson loops with increasing area (see below).
It seems therefore, that the potential screening effects due to light
dynamical fermions are capable of being reproduced by these first order
estimates based on quenched configurations.


\section{Extensions}
\label{Sec:Extensions}

\subsection{Naive Improvement}

The results of the last section are quite encouraging.  Approximating
$T$
with a $1\times 1$ plaquette is seen to shift significantly
the quenched results towards the full theory results.
The absolute measure of how well we have actually done was defined in
the discussion following 
(\ref{Eqn:deltadeltaexpanded}) and is given
by the corresponding values of $\sigma^2(\Delta)$.
For the $32^2$ lattice at $\beta=2.50$, $\kappa=0.26$, we find
$\sigma^2(T) = 99.0(2)$,  while 
$\sigma^2(T+\delta\beta \cW_{1\times 1}) = 3.1(7)$.
These values allow us to predict acceptances
according to (\ref{Eqn:Erfc2}) for algorithms which first generate
trial configurations, then accept/reject those configurations
according to the change in $T$ or $T+\delta\beta \cW_{1\times 1}$
respectively.
The value of $\sigma^2(T) = 99.0(2)$
implies that an algorithm which uses
the pure gauge term of the Schwinger model at unshifted $\beta$
to generate trial configurations will have
an acceptance of $2\times 10^{-12}$ (${\rm erfc}(0.5 \sqrt{99.0})$).
The value $\sigma^2(T+\delta\beta \cW_{1\times 1}) = 3.1(7)$,
on the other hand, implies that an algorithm which uses 
the pure gauge term at a properly tuned shifted $\beta$ will
have an acceptance of 0.2.
For the Schwinger model, at the parameters we have analyzed, 
the $1\times 1$ loop is therefore seen to be a reasonable 
approximation
to the fermion determinant, since it produces a finite (non zero)
value of acceptance.  It is not perfect, however, since acceptance
is still quite small.

We wish to consider now whether, and to what extent,  the addition of further loops
to the approximate action can improve the situation.
For this and all remaining analysis in this paper, 
we will concentrate on the effects of approximate actions
on loop operators only.  The single loop analysis just described
shows that the loop expectation values are the most sensitive to changes in
action.
The effective mass is rather insensitive, and in any case,
is already quite well described by the addition of just the $1\times1$
plaquette term in the approximate action.

\begin{figure}[t]
\medskip
\centerline{\psfig{figure=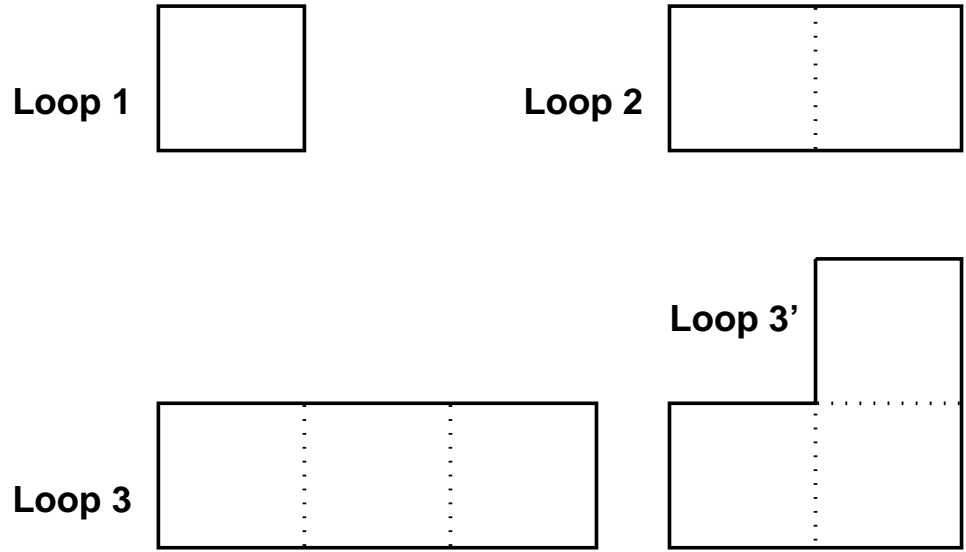,width=5.0in,angle=270}}
\medskip
\caption{Loops included in the first attempt to improve on
the quenched approximation for the Schwinger model.}
\label{Fig:Loops}
\end{figure}

The first addition which we considered involves an approximate action
where we included loops containing up to three plaquettes.
The
loops involved are the $1\times 1$, 
$1\times 2$, and $1\times 3$ rectangles 
(denoted by 1, 2, and 3 for short), 
together with a \lq chair\rq{} shaped loop of length 8 (denoted
by $3'$). These four loops are shown in Figure \ref{Fig:Loops} and
are some, but not all, of the simple loops of
perimeter length 8 or less.
An immediate problem with this loop expansion  is that the
loop possibilities expand very rapidly as the loop length
increases. Even at this simple expansion level, we are already
ignoring some non-trivial length 8 loops (for example, the $1\times 1$
squared loop operator).

The target full fermion action of our analysis has $\beta=2.50$ and
$\kappa = 0.26$.  Expressed in terms of Gram-Schmidt orthogonalized
loop operators, the generic approximation of $T$  is now
\begin{equation}
T = a_1 \hat\cW_1 + a_2 \hat\cW_2 + a_3 \hat\cW_3 + a_{3'} \hat\cW_{3'} \, .
\end{equation}
Initially we have no information about the values of the coefficients
$a_i$, so we set them to zero and simulate with an approximate action
which contains only the pure gauge part of the Schwinger action with
$\beta = 2.50$ (\ie we simulate with the unshifted quenched version of
the Schwinger action).  The ensemble so generated allows us to make new
estimates of the coefficients $a_i$, which can be used to generate a new
approximation.  We repeated this process four times in all.  Table
\ref{Tab:Approximations} summarizes the sequence followed.  

Ensembles
of 2000 configurations were generated for each approximation, and
the approximate action coefficients for the next approximation were
calculated on these ensembles using the formulae of Section
\ref{Sec:WilsonLoops}.  The results achieved are given in 
Table \ref{Tab:AValues}.
Approximation 0 in a quenched simulation with
$\beta=2.50$.  The value which results for $a_1^{(0)}$ from this
approximation is $0.212(10)$
which is exactly the value of $-\delta\beta$ determined in the last
section.  Here however, we have specified that the true
action has $\beta=2.50$.
Thus we interpret $a_1^{(0)}$ as defining the best quenched
approximation coupling shift to use to simulate the full theory with
$\beta = 2.50$.  This best quenched theory has $\beta' = 2.50 +
0.212(10)$ and defines our second approximation (approximation 1 in
Table \ref{Tab:Approximations}).  Each approximation of course
generates a different ensemble, and the values for the approximate
action coefficients $a_i$ therefore change from approximation to
approximation.  This is clear also in the table.  But a very nice 
feature to note is the fundamental stability of the procedure we have
adopted.  The coefficients which result after approximation 3 are, within errors,
the same as the coefficients after approximation 0.

For the highest approximation attempted
(approximation 3), we also list the corresponding root variance
$\sigma(\hat{\cW}_i)$ and the product $a_i\sigma(\hat\cW_i)$.  
We have, when the coefficients $a_i$ are tuned according to
\ref{Eqn:a_i},
that
\begin{equation}
\sigma^2(T - \sum_i a_i \hat\cW_i) 
= \sigma^2(T) - \sum_i a_i^2 \sigma^2(\hat\cW_i)
\, .
\end{equation}
Thus, $a_i\sigma(\hat\cW_i)$ is a measure of the importance of
a particular loop operator in approximating $T$.   If the addition
of loops of increasing size generated a rapidily converging
approximation
to the true action, we would expect $a_i\sigma(\hat\cW_i)$
to decrease rapidly as the loop size increases.  
It is clear
from the table, however, that this is not the case for the Schwinger model.
The contributions of two and three plaquette loops are not significantly
smaller than the contribution of the single plaquette loop.

\begin{figure}[t]
\medskip
\centerline{\psfig{figure=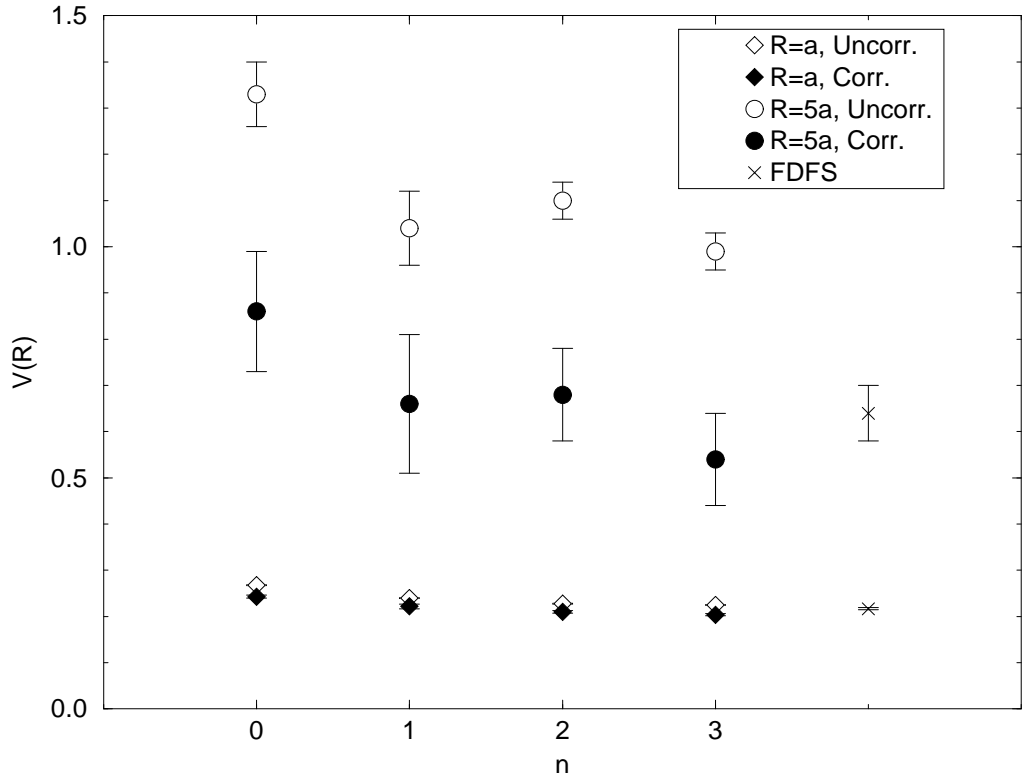,width=5.0in,angle=-90}}
\medskip
\caption{The static potential in lattice units using different orders
of approximation. Open symbols denote raw measurements at a given order,
while full symbols denote deficit corrected results.
The corresponding full dynamical simulation results are indicated by the
crosses at the right of the diagram.}
\label{Fig:FixedRPotential}
\end{figure}

In Figure \ref{Fig:FixedRPotential}, we show the static potential, at
fixed distance in lattice units (for $R=a$ and $R=5a$), versus the
approximation number, $n$.  We show both the raw measurements made using
each set of configurations, and the deficit-corrected estimates given by
(\ref{Eqn:FCumExpans}) as described above.  The corresponding dynamical
fermion result (2,000 configurations) is also shown for each inter-quark
distance.  As expected, we see that adding loops of increasing
size has very little effect on the 
short distance potential at $R=a$. However, at $R=5a$ there is an effect,
and a decreasing trend is
evident in the (uncorrected) data obtained as larger loops are added to 
approximate action.  This decrease is quite slow unfortunately, and
confirms
that
the systematic addition of loops of increasing size produces only
a slowly converging approximation to the true action. 

One significant positive observation can made from Figure
\ref{Fig:FixedRPotential}.  This observation, anticipated in the
previous section, is that at each approximation order, $n>0$, the
deficit-corrected estimate given by (\ref{Eqn:FCumExpans}) is compatible
with the dynamical fermion result. At first sight, it may therefore seem
that there is no need to go beyond $n=1$, \ie to use other than quenched
data to make properly corrected estimates of the full theory. However,
as emphasized in Section \ref{Sec:Tuning}, one can only be confident of
such cumulant expansion-based estimates when the shifts involved are
known to be small. This will only be so when the corrections are based
on a suitably accurate approximate action.

Note finally that the corrections are not simply shifts in $\beta$,
but are non-trivial effects of dynamical fermions. One may think of the
estimates as a form of re-weighting in which the approximate vacuum used
becomes closer to the true vacuum as more loops are added. However,
what is very clear is that the addition of small loops only in the
approximate action improves physics only at the scales appropriate for
small loops.  Thus the static potential at short range is almost
perfectly determined by the approximate action.  At longer range,
however, the approximate action is not very successful in improving the
quenched result.

\subsection{Selective Improvement}

This analysis suggests that an approximate action based
on a systematic expansion of $T$ in terms of loops of increasing
size is going to require the inclusion of a very large
number before convergence is achieved.   However, the impression
already gained is that, once  a loop of a particular size
is added to the approximate action, the addition of more loops of a
similar size is not particularly productive. This suggests an
alternative approach of adding loops selectively rather than
systematically. The aim would be  to include typical loops of different
sizes rather than all loops up to a given size. As an experiment, we
chose to consider approximate actions with $1\times 1$, $3\times 3$ and
$5\times 5$ loops included in turn.  For simplicity,
tuned coefficients defining all these
actions were determined on a single quenched, $\beta=2.50$ ensemble.

The results achieved by these new approximate actions are given in
Table \ref{Tab:corrected}.
The leftmost column shows the values for various loop expectations
from a full dynamical fermion simulation.  The remaining
columns gives the values achieved by the various different
approximate actions. 
What we see is that the expectation value achieved for a
loop in a simulation which includes that loop in the approximate
action correctly matches its full dynamical simulation value.
This is in good agreement with
our analysis of Section \ref{Sec:Tuning} where we argued that our approximate
action tuning conditions are such as to match loops included in the
approximate action to their correct values in the true action.
Loops not explicitly included in the approximate action are not so well
tracked.  There is however considerable improvement as we add more loops
to
the approximate action.  For example, the $3\times 8$ loop changes
from value 3.7 when only
the $1\times 1$ term is included in the approximate action to the value 10.5
when $1\times 1$, $3\times 3$, and $5\times 5$ loops are included.
This is to be compared with
its exact full theory value of 14.4.

A check on the results achieved is to execute an iteration of the
tuning algorithm to see if our results are sensitive to the choice of
coefficients in the $1\times 1 + 3\times 3 + 5\times 5$ approximate
action used.  
To execute this iteration, we used our $1\times 1 + 3\times 3 + 5\times
5$ 
approximate action to generate a new ensemble on which we recalculated
the coefficients $a_{1\times 1}, a_{3\times 3}$ and $a_{5\times 5}$.
Table \ref{Tab:loops2} shows the results of this retuning.
Again we find that the results are quite stable.  In the Schwinger
model, a set of quenched configurations seems sufficient to calculate
the approximate action coefficients.

\subsection{Optimized Improvement of The Effective Action}

\begin{figure}[t]
\medskip
\centerline{\psfig{figure=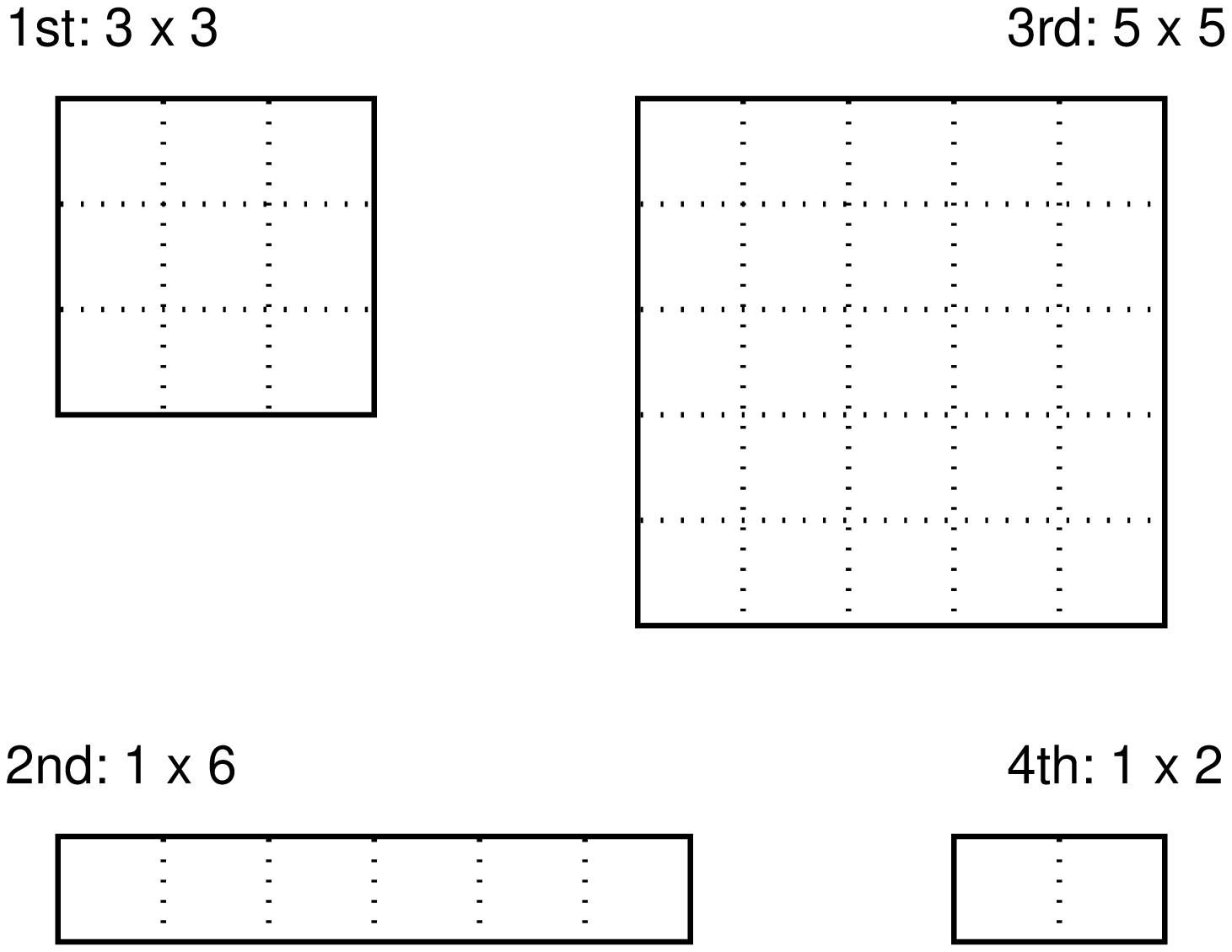,width=5.0in,angle=270}}
\medskip
\caption{Loops which give the optimum approximation to
the true Schwinger model dynamical fermion action.}
\label{Fig:OptimumLoops}
\end{figure}

As a final test of the method, we attempt a very direct approach
to defining the effective action by using an ensemble of configurations
generated using the full fermion algorithm.
The ensemble is for a $16^2$
lattice
at $\beta=2.50$ and $\kappa=0.26$.  In all we generated 1,000
configurations, and
systematically considered a large number of different possible
approximations for $T$ of the form
\begin{equation}
T \approx \sum_i a_i \cW_i
\end{equation}
where the index $i$ here runs over some selected subset of the Wilson
loop
operators which we have calculated.  For example, one possible
approximation
to consider is
$T\approx a_{2\times 2}\cW_{2\times 2}+a_{1\times 3}\cW_{1\times 3}$.
For each possible approximation, we choose values of $a_i$ which
minimize
$\sigma^2(T-\sum_i a_i \cW_i)$ for that approximation, and then
determine the
approximate action which generates the minimum value for $\sigma^2$.
The results of this minimization are listed in Table \ref{Tab:LoopExpt},
and the optimum loop choice is shown in Figure \ref{Fig:OptimumLoops}.
The first line of the table gives the value of $\sigma^2(T)$ with
no approximation attempted.  The second line indicates that the
best single loop approximation to $T$ is obtained when the loop
is the $3\times 3$ loop.  The best two loop approximation
is obtained by using the $3\times 3$ and $1\times 6$ loops, etc.
The measure of the quality of a given approximation is the value of
$\sigma^2(T-\sum_i a_i \cW_i)$ listed in the final column of this table.
The best one loop approximation reduces $\sigma^2$ from $2.69$ to
$1.02$.  Adding more loops reduces $\sigma^2$ further, but not
at the same rate as adding only a single loop.

\subsection{Stability}

In all cases considered so far, we have tuned the effective action
by minimizing $\sigma^2$.  In Section \ref{Sec:Tuning} we proposed
three different tuning conditions.  These all agree to lowest order in
the
cumulant expansion.  They differ at higher order, however.  To test
whether the differences between the three tuning conditions are
significant, we compared the coefficients for the effective actions
tuned according to (\ref{Eqn:MaximizeOverlap}),
(\ref{Eqn:MatchCondition}) and (\ref{Eqn:MaximizeAccept}).
The results of this comparison are shown in table \ref{Tab:TuningShifts}.
The values determined by minimizing $\sigma^2$
for the coefficients of the $3\times 3$,
$1\times 6$, $5\times 5$ and $1\times 2$ loops in the best four loop
approximation for $T$ are given in the second column of the table.
If instead, we tune to match the expectations of the four loops
in the approximate action between the true and approximate measures,
then the values of the coefficients are shifted by the amounts listed
in the third column of the table.  If we tune to maximize acceptance,
then the shifts necessary are given in the final column of the table.
In all cases the shifts are very small, compatible with zero
at this level of statistics.  This is indicative that
we have found an approximate action which is sufficiently close
to the true action that all the tuning conditions which we have
considered are self-consistent.

Having arrived at a tuned approximate action (\eg coefficients
given in Table \ref{Tab:TuningShifts}), we have
made preliminary tests of an exact algorithm as described in
Section \ref{SubSec:MaximizeAccept}.
A similar exact algorithm was introduced in the context of the
bosonic algorithm \cite{Luscher} by Peardon \cite{Peardon} and further
developed by others \cite{deForc}.
These tests have shown that correct results can indeed be obtained
with an experimental acceptance as predicted in Section
\ref{SubSec:MaximizeAccept}.
However, careful tuning is required
when applying the algorithm to large systems. The approximate
action must yield $\sigma^2(\Delta_{at}$ not too different from $1$ so that
reasonable acceptances are obtained.


\section{Application to QCD}
\label{Sec:QCD}

So far, the main numerical results in this paper have concerned the
application of approximate actions to the Schwinger model. We chose
this model because it allowed useful comparisons between the ensembles
generated by an exact full fermion algorithm and those generated by a
variety of different approximate actions in reasonable computational
time.  However, our main interest in the application of approximate
actions is in lattice QCD. A complete repetition of the analysis so far
described for lattice QCD is a considerable computational undertaking,
and beyond the scope of this exploratory paper. Nevertheless, it has been
possible to gain some indication of both the potential of the method and
the practical problems which still remain for lattice QCD.

The exploratory simulation which we have carried out is for full
QCD on a small lattice.  The true action which we considered
was the standard
single plaquette pure gauge action with coupling $\beta$,
combined with a two-flavour Wilson fermion action with hopping
parameter $\kappa$.  The lattice size adopted was $6^4$ with
$\beta = 5.7$ and $\kappa = 0.16$.  This is obviously a very small
lattice with very heavy quarks.  The basic goal
was to determine the extent to which it is possible
to approximate the true action for this theory with an action made
up from a finite set of Wilson loops.

The direct approach to analyzing this question follows exactly
the strategy adopted at the end of the last section, and
is simply to
generate an ensemble of configurations using the full
QCD action, to evaluate $T$
and various different loops
on these configurations, and then to determine the set of loops
which best approximates $T$. This is the approach we have
adopted.
We used hybrid molecular dynamics to generate configurations.
The ensemble analyzed contained 516 configurations.  Each
configuration was separated by 10 trajectories of length 0.5.
$T$ was calculated as for the Schwinger
model on each configuration.
We also calculated a variety of different Wilson loop operators
on each configuration.  These loops were chosen to be representative
rather than complete.  Four dimensions allow many more possible
loop configurations than two dimensions, and an exhaustive study
of all the loop possibilities is not justified in the present
instance where we are working with a very unphysical set of
lattice parameters. Our goal is simply to explore the possibilities
rather than to present a definitive calculation.

Following our experience with the Schwinger model, we looked at
a variety of different loop shapes and sizes.  These
included two, three, and four dimensional loops, with total length
varying from 4 to 12 lattice spacings.  The only non-trivial loop
of length 4, of course, is the pure gauge $1\times 1$ plaquette action,
while the
length 12 loops considered included $5\times 1$, $4 \times 2$, and
$3\times 3$ 2-dimensional loops together with more complicated
three and four dimensional constructions.  A particular loop
construction is defined by specifying the steps in the various
possible directions which a link in that loop can represent.
Thus the specification $\{12\bar{1}\bar{2}\}$ denotes a loop generated
by first stepping in the direction $1$ with positive sense,
then in direction
$2$ with positive sense, then in direction $1$ with negative
sense, and
finally in direction $2$ with negative sense.  This set of four steps
brings
the loop operator back to its starting site. We can then
take the trace in the usual way.  The digits $1$ and $2$ can represent
any two distinct directions in the four dimensional lattice.  An
unbarred digit
defines a step in the positive direction along the corresponding
axis, a barred digit defines a step in the negative direction.
The specification implicitly assumes a sum over all independent
orientations and starting locations of the loop.  For example, the
specification we have just written represents the simple
$1\times 1$ plaquette summed over all orientations and lattice sites,
and is therefore the standard  one plaquette pure gauge action term.

To test how well the true action can be approximated, we
then seek to minimize $\sigma^2(T - \sum_i a_i \cW_i)$ over the
different loops considered.
In all, we calculated over 40 loops. 
This allows us to consider a very large number of different
approximate
actions.  These possibilitys include actions with a single loop include,
actions with two loops included etc.

Table \ref{Tab:QCDData} shows some of the results we achieved. 
The first
column lists the variety of loops which we considered.  The
order in which these loops are listed represents
their relative importance in reducing fluctuations
in the quantity $\sigma^2(\Delta)$ and is determined only after all analysis
is complete.
The second column lists the minimum achieved for $\sigma^2(\Delta)$ when just
a single loop is included in the approximate action.  This minimum,
of course, varies depending on which particular loop is considered
as the single term approximate action, and the second column
shows this variation. The raw value of the variance in
$T$ is listed at the end of the table for comparison purposes.

It is immediately noteworthy that there
is a large variation in the values achieved for $\sigma^2(\Delta)$, indicating
that there is a large variation in the effectiveness with which
different  loops can individually approximate $T$.  The effective
loops tend to be loops with four dimensional structure,
and it is quite striking that the simplest loop of all, the single
plaquette loop (defined as $\{12\bar{1}\bar{2}\}$) is actually quite
unimportant. The best match between true and approximate
actions is given by the loop which is
the first entry in the table.  This loop, on its own,
reduces the fluctuations in $S_t$ from a value of $295 (20)$
to a value $8.7 (4.4)$.
This is a remarkable 97\% reduction
in the raw fluctuations of $S_t$.
The loop which does this is
not a simple loop but rather an oddly shaped four dimensional
construction which has total length 12.  We have not been able to
identify any remarkable features about this particular loop
relative to all the other loops which we considered.
Other similarly shaped loops are seen to reduce the fluctuations in
$S_t$ almost as much, so it is reasonable to imagine that
the important feature is the four dimensional nature of the
loop and its approximate size rather than the exact details
of how the loop is generated.

The determination of the best single loop approximate action
is only the first analysis step.  The next is to determine
the best two loop approximate action.
Our approach is exactly as before.  We minimize $\sigma^2(\Delta)$ for
all two loop approximate actions.  In fact, we only need to minimize
$\sigma^2(\Delta)$ over loops with $1$ and $n$ where $n$ is any of the other
available loops.  This minimization determines that the approximate
action which uses the loops placed  1 and 2 in the table
is the best two loop approximate
action.  The final column of Table \ref{Tab:QCDData} indicates the
value which results for $\sigma^2(\Delta)$ from this minimization.  The minimization
for higher numbers of terms in the effective action can now proceed
by iteration.  The final result we find is that the best $n$ loop
approximate action
is the action which contains loops $1, 2, \ldots, n$.
The values which result for $\sigma^2(\Delta)$ with this best $n$ term approximate
action are all given by the final column of Table \ref{Tab:QCDData}.

The second interesting feature of this exploratory study is that
the addition of extra loops beyond the first loop in the approximate
action is not particularly helpful.  The approximate action
with a single loop reduces the raw fluctuations in $S_t$ from $295 (20)$
to $8.7 (4.4)$.
Adding a second and third
term reduces these fluctuations only to $7.2 (4.1)$ and $5.95 (4.00)$
respectively.
These shifts
are very much smaller  than that achieved by a single term alone.

As even more loops are added, the error in $\sigma^2(\Delta)$ becomes greater
than its actual value, and we conclude that we have insufficient
data to calculate reliably the best approximate action for more than
a small number of terms, so the actual ordering after the
first three or four terms should be considered suspect. However, a
significant observation which can be made is that the $1\times 1$
plaquette loop contribution comes at position 35, which
is very far down the list.  We conclude therefore
that the pure gauge plaquette action is actually a rather poor
approximant of the the full fermion action.


\section{Conclusions}
\label{Sec:Conclusions}

We have proposed a general framework for the construction and evaluation
of approximate  actions for lattice QCD. We have demonstrated how the
coefficients in the simplest form of such actions can be evaluated
directly from a true action simulation or by using an
iterative procedure which begins with an approximate
action simulation, and never requires simulation with the true action.  We have
also shown that the results of the iterative procedure are quite stable,
and converge very quickly.

As a first application, we have tested approximate actions
based on the systematic expansion of the trace of the log of the fermion
matrix in Wilson loops of increasing size
as originally proposed in \cite{SW1}.  This expansion works well for
observables sensitive to short range physics. Using the Schwinger model,
we were able to test longer range observables where it becomes clear that an
impractically large number of loops will be required to study physics
over a range of scales using small lattices spacings.

The systematic error correction estimate proposed by \cite{SW1}
is shown to give reliable estimates of observables based on
approximate action simulations. The better the approximate action, the
more reliable these \lq deficit\rq{} estimates become. However,
since these
estimated corrections require a good, unbiased,  stochastic estimate of
$T$, the trace of the log of the fermion  matrix, their practical value is dependent on
further progress in developing efficient calculational methods for this
stochastic estimation. The computational methods which we have used here to evaluate
$T$ are quite demanding.  Typically they require work
equivalent to
that needed to evaluate all meson and hadron propagators at a number
of different hopping parameters on a given configuration.

Having established  the above behaviour, we have gone on to show
that the greatest determinant of the quality of an effective action is
not the number of loops included,
but the  choice of scales represented by the loops which are included.
We have found dramatic evidence of this both for the two-dimensional test
model and for QCD itself. The systematic matching and tuning methods have
allowed us to propose approximate actions significantly closer to the
true QCD action than the quenched approximation or, indeed, of an approximation
based on the ordered inclusion of loops of increasing size.

We presented two methods for capitalising on approximate actions
in lattice simulation.
The first, as originally proposed in \cite{SW1},
is to simulate with the approximate action and to make (presumably) reliable
estimates of the small corrections to desired observables.
The challenge to progress in this strategy is to improve stochastic
evaluation methods for $T$.
The second method described is to use the approximate action to generate candidate
update configurations within an exact algorithm, and we proved by
construction that
such an exact simulation was feasible in the Schwinger model example.
The challenge to the construction of such an exact algorithm in QCD
is to implement the fine tuning required to achieve viable acceptance
rates.  It is perhaps too optimistic to imagine that this fine tuning
can be achieved by considering actions made of Wilson loop operators
only.  The QCD test described here suggests that the return achieved
in adding more Wilson loops to the approximate action becomes very poor
after the first few loops.  On the other hand, the tuning methods we
have defined are completely general, and can be applied
to other candidates for approximate actions.  In particular, a
combination approximate action including a small number of Wilson
loop operators and a multiboson
L\"{u}scher \cite{Luscher} term seems to us to be an interesting
possibility as a tunable candidate approximate action.

While awaiting these developments, an immediate and exciting
application of the methods we have described
is to perform relatively high statistics measurements using
the approximate action directly. The strategy here is to make low
statistics measurements of the quantity of interest using the full
theory. These can then be used to help tune an  approximate action to
reproduce this measurement approximately (and  other more accessible
ones to higher levels of accuracy). The approximate action can then be
used to give high statistics estimates of the quantity of interest (\eg
glueball mass). 

%

\section{Acknowledgements}

One of us (J.S.) would like to thank D.\ Weingarten for much discussion
on the subject matter of this paper.  We would also like to thank
The Hitachi Dublin Laboratory for providing computer resources which
we used to perform some of the calculations described here.


\begin{table}[t]
\caption{Sequence of approximations to $T$ used to test the effects
of adding loops of increasing size to generate an approximate
Schwinger model action.  The loops included in these
approximations are defined in Figure \ref{Fig:Loops}.  The superscripts $(n)$
on the coefficients $a_i^{(n)}$ 
indicate the approximation used to calculate those coefficients.
Note that, in all cases, approximation $k$ uses coefficients
determined
by simulating with approximation $k-1$.}
\label{Tab:Approximations}
\begin{tabular}{rl}
Approximation & Form of $T$ used \\
\hline
0 & $ T = 0$ \\
1 & $ T = a_1^{(0)} \hat\cW_1 $ \\
2 & $ T = a_1^{(1)} \hat\cW_1 + a_2^{(1)} \hat\cW_2 $ \\
3 & $ T = a_1^{(2)} \hat\cW_1 + a_2^{(2)} \hat\cW_2 
          + a_3^{(2)} \hat\cW_3 + a_{3'}^{(2)} \hat\cW_{3'} $\\
\end{tabular}
\end{table}


\begin{table}[t]
\caption{Numerical values for the coefficients $a_i$
in the expansion of $T=\sum_i a_i \hat\cW_i$ and
their relative contributions (see text).}
\label{Tab:AValues}
\begin{tabular}{l|llllll}
$\phantom{\cW_t}$         & $a_i^{(0)}$           & $a_i^{(1)}$
& $a_i^{(2)}$           & $a_i^{(3)}$
& $\sigma^{(3)}(\hat\cW_i)$      & $a_i^{(3)}\sigma_t^{(3)}(\hat\cW_i)$     \\
\hline
$\hat{\cW_1}$     &$0.212(10)$    &$0.208(10)$    &$0.197(11)$    &$0.222(10)$
&$10.2(2)$      &$2.26(11)$     \\
$\hat{\cW_2}$     &$0.085(7)$     &$0.083(7)$     &$0.073(7)$     &$0.083(7)$
&$13.8(3)$      &$1.15(10)$     \\
$\hat{\cW_{3}}$   &$0.041(6)$     &$0.018(11)$    &$0.034(7)$     &$0.048(8)$
&$12.7(2)$      &$0.61(10)$     \\
$\hat{\cW_{3'}}$  &$0.025(4)$     &$0.026(5)$     &$0.034(5)$     &$0.029(4)$
&$23.9(4)$      &$0.69(10)$     \\
\end{tabular}
\end{table}


\begin{table}[t]
\caption{Rectangular Wilson loop measurements made on configurations generated
by full dynamical fermion simulation (true), by an approximate action with
$1\times 1$ loop added (column 2), with $1\times 1$ and $3\times 3$
loops
added (column 3), and with $1\times 1$, $3\times 3$, and $5\times 5$
loops
added (column 4).  In a given column, expectations
of loops included in the corresponding approximate action are
marked with an asterisk.
}
\label{Tab:corrected}
\begin{tabular}{l|rrrr}
        & true          & $1\times 1$           & $+3\times 3$  & $+5\times 5$           \\
\hline
$1\times 1$     & 805.1(3)      & 805.5(2)*     & 804.9(2)*     & 804.4(2)*     \\
$2\times 2$     & 415.7(9)      & 391.8(7)      & 410.4(8)      & 408.2(8)      \\
$3\times 3$     & 158.9(11)     & 118.9(9)      & 162.4(10)*    & 160.6(10)*    \\
$4\times 4$     & 51.2(10)      & 23.7(7)       & 47.0(8)       & 46.7(9)       \\
$5\times 5$     & 15.1(8)       & 3.8(7)        & 8.9(7)        & 13.8(6)*      \\
$1\times 2$     & 640.2(6)      & 633.5(4)      & 637.6(4)      & 636.4(4)      \\
$1\times 3$     & 512.2(8)      & 498.2(5)      & 507.5(5)      & 506.3(6)      \\
$1\times 4$     & 411.2(10)     & 392.0(6)      & 403.9(6)      & 402.9(7)      \\
$1\times 5$     & 330.9(11)     & 308.5(7)      & 321.5(7)      & 320.8(8)      \\
$1\times 8$     & 172.9(11)     & 150.4(7)      & 162.0(8)      & 162.3(9)      \\
$2\times 6$     & 88.8(9)       & 58.4(6)       & 77.9(7)       & 78.0(8)       \\
$3\times 8$     & 14.4(6)       & 3.7(5)        & 10.0(5)       & 10.5(6)       \\
\end{tabular}
\end{table}


\begin{table}[t]
\caption{
Comparison of Wilson loop expectations calculated
with a true full dynamical simulation, and with 
approximate actions containing $1\times 1$, $3\times 3$
and $5\times 5$ loops.  Column 2 lists the loop expectations
determined with approximate action coefficients tuned on
a quenched ensemble.  The approximate action coefficients
in column 4 are tuned on an ensemble generated from the approximate action
of column 2, and represent therefore a retuning of the coefficients
generating column 2.  
Columns 3 and 5 list the corresponding deficit corrected
loop expectations for each approximate action.}
\label{Tab:loops2}
\begin{tabular}{l|ddddd}
loop & true action  & quenched tuning & quenched tuning & retuned & retuned  \\
     &              &                 & (corrected)     &                 & (corrected) \\
\hline
$1\times 1$     & 201.5(02)      & 201.2(02)      & 201.2(02)      & 201.5(01)      & 201.5(01)  \\
$2\times 2$     & 105.2(06)      & 101.8(08)      & 104.4(11)     & 103.3(05)      & 104.3(08)  \\
$3\times 3$     & 40.5(07)       & 39.8(07)       & 39.8(07)       & 42.6(08)       & 42.6(08)   \\
$4\times 4$     & 12.9(06)       & 12.4(05)       & 13.1(10)      & 13.7(03)       & 14.1(08)   \\
$5\times 5$     & 3.6(04)        & 3.7(04)        & 3.7(04)        & 5.1(04)        & 5.1(04) \\
$1\times 2$     & 160.7(04)      & 159.2(04)      & 160.2(06)      & 159.8(02)      & 160.2(03)  \\
$1\times 3$     & 128.9(05)      & 126.5(06)      & 128.5(07)      & 127.5(02)      & 128.5(04)  \\
$1\times 4$     & 103.4(06)      & 100.5(07)      & 103.2(09)      & 101.6(03)      & 102.9(06)  \\
$1\times 5$     & 82.9(06)       & 79.8(08)       & 83.2(12)      & 81.0(03)       & 82.4(08)   \\
$1\times 8$     & 42.5(07)       & 39.5(08)       & 43.2(14)      & 41.1(05)       & 42.9(12)  \\
$2\times 6$     & 22.2(06)       & 19.2(07)       & 21.0(13)      & 20.5(03)       & 21.6(06)   \\
$3\times 8$     & 3.8(03)        & 2.8(04)        & 2.0(08)        & 3.6(04)        & 3.3(08)    \\
$5\times 7$     & 1.4(03)        & 0.8(02)        & 2.0(06)        & 0.9(02)        & 1.2(06)    \\
$5\times 8$     & 0.4(02)        & 0.4(02)        & 0.4(05)        & -0.1(03)       & 1.2(05)  \\
\end{tabular}
\end{table}


\begin{table}[t]
\caption{
The residual variance corresponding to an approximate action
incorporating the Wilson loop terms shown. The variance is a measure
of the quality of the approximate action.
}
\label{Tab:LoopExpt}
\begin{tabular}{rrd}
Terms &  Loops & $\sigma^2(T - \sum_i a_i \cW_i)$ \\
\hline
0 &               & 2.69 (27) \\
1 &   $3\times 3$ & 1.02 (09) \\
2 & $+ 1\times 6$ & 0.61 (07) \\
3 & $+ 5\times 5$ & 0.59 (07) \\
4 & $+ 1\times 2$ & 0.31 (05) \\
\end{tabular}
\end{table}


\begin{table}[t]
\caption{
Effect of tuning condition choice on the loop coefficients in a tuned
approximate action. 
Loop coefficient values as determined by minimizing 
$\sigma^2$ are given in the second column.  The shifts in these values
when  higher order tuning conditions are incorporated are given in the
remaining columns.}
\label{Tab:TuningShifts}
\begin{tabular}{r|ddd}
coefficient & $\sigma^2$ & Operators & Acceptance \\
     & Minimized & Matched & Minimized \\
\hline
$a_{3\times 3}$  & 0.2150 (12) & -0.013 (45) & -0.026 (94) \\
$a_{6\times 1}$  & 0.0325 (18) & -0.001 (5) & -0.001 (10) \\
$a_{5\times 5}$  & 0.0105 (20) & -0.004 (5) & -0.006 (8) \\
$a_{1\times 2}$  & 0.0109 (22) & -0.003 (9) & -0.004 (18) \\
\end{tabular}
\end{table}


\begin{table}[t]
\caption{Approximate action optimization data for the lattice QCD test
described in the text.
Wilson loop operators are defined in column 2.  
Column 3 lists the 
value of $\sigma^2(T - a_i \cW_i)$ (no sum on $i$) for the
loop operator defined in that row. 
Column 4 lists the value
of $\sigma^2(T - \sum_{k=1}^i a_i \cW_i)$ where the 
sum includes all the loop operators at that row or higher in the table.
The order in which loop operators are listed is the order which
minimizes $\sigma^2(T - \sum_{k=1}^i a_i \cW_i)$ when at most $i$ loops
are included.   Note that we present information only on those
loops which generate better approximate actions than the single 
plaquette loop, which appears at position 35.
}
\label{Tab:QCDData}
\begin{tabular}{rlrr}
Order & Specification & $\sigma^2[i]$ & $\sigma^2[1+\cdots+i]$\\
\hline
 1 &$\{11234\bar{3}\bar{1}\bar{1}\bar{4}\bar{2}\}$
    &$  8.7 \; (\pm 4.4) $&$8.7 \; (\pm 4.4) $ \\
 2 &$\{111222\bar{1}\bar{1}\bar{1}\bar{2}\bar{2}\bar{2}\}$
    &$ 82.7 \; (\pm 9.2) $&$7.2 \; (\pm 4.1) $ \\
 3 &$\{123413\bar{1}\bar{4}\bar{3}\bar{2}\bar{1}\bar{3}\}$
    &$ 12.6 \; (\pm 4.4) $&$6.0 \; (\pm 4.0) $ \\
 4 &$\{123\bar{1}\bar{2}\bar{3}\}$
    &$ 14.4 \; (\pm 5.0) $&$5.8 \; (\pm 3.9) $ \\
 5 &$\{1213\bar{1}\bar{2}\bar{1}\bar{3}\}$
    &$  9.3 \; (\pm 4.5) $&$5.6 \; (\pm 3.9) $ \\
 6 &$\{12\bar{1}\bar{2}34\bar{3}\bar{4}\}$
    &$ 22.5 \; (\pm 5.6) $&$5.3 \; (\pm 3.8) $ \\
 7 &$\{1112\bar{1}\bar{1}\bar{1}\bar{2}\}$
    &$ 19.0 \; (\pm 4.6) $&$5.1 \; (\pm 3.7) $ \\
 8 &$\{112\bar{1}\bar{1}\bar{2}334\bar{3}\bar{3}\bar{4}\}$
    &$ 17.1 \; (\pm 4.8) $&$5.0 \; (\pm 3.8) $ \\
 9 &$\{1233\bar{2}\bar{1}\bar{3}\bar{3}\}$
    &$  9.5 \; (\pm 4.6) $&$4.8 \; (\pm 3.7) $ \\
 10&$\{111122\bar{1}\bar{1}\bar{1}\bar{1}\bar{2}\bar{2}\}$
    &$ 60.1 \; (\pm 8.4) $&$4.6 \; (\pm 3.6) $ \\
 11&$\{112334\bar{3}\bar{3}\bar{1}\bar{1}\bar{4}\bar{2}\}$
    &$  9.6 \; (\pm 4.6) $&$4.4 \; (\pm 3.7) $ \\
 12&$\{11223\bar{2}\bar{2}\bar{1}\bar{1}\bar{3}\}$
    &$ 14.4 \; (\pm 4.6) $&$4.0 \; (\pm 3.9) $ \\
 13&$\{112\bar{1}\bar{1}\bar{2}\}$
    &$ 15.2 \; (\pm 4.5) $&$3.6 \; (\pm 3.7) $ \\
 14&$\{12\bar{1}\bar{2}13\bar{1}\bar{3}\}$
    &$ 50.1 \; (\pm 7.5) $&$3.5 \; (\pm 3.8) $ \\
 15&$\{12\bar{1}31\bar{2}\bar{1}\bar{3}\}$
    &$ 20.0 \; (\pm 5.2) $&$3.2 \; (\pm 3.9) $ \\
 16&$\{122\bar{1}\bar{2}\bar{2}13\bar{1}\bar{3}\}$
    &$ 33.1 \; (\pm 5.6) $&$3.1 \; (\pm 3.8) $ \\
 17&$\{11233\bar{1}\bar{1}\bar{2}\bar{3}\bar{3}\}$
    &$ 17.7 \; (\pm 4.6) $&$3.0 \; (\pm 3.8) $ \\
 18&$\{112233\bar{2}\bar{2}\bar{1}\bar{1}\bar{3}\bar{3}\}$
    &$ 23.0 \; (\pm 4.7) $&$2.7 \; (\pm 3.8) $ \\
 19&$\{11233\bar{2}\bar{1}\bar{1}\bar{3}\bar{3}\}$
    &$ 14.9 \; (\pm 4.6) $&$2.6 \; (\pm 3.7) $ \\
 20&$\{122\bar{1}331\bar{2}\bar{2}\bar{1}\bar{3}\bar{3}\}$
    &$ 24.3 \; (\pm 5.2) $&$2.6 \; (\pm 3.7) $ \\
 21&$\{123\bar{2}\bar{1}\bar{3}\}$
    &$ 14.2 \; (\pm 5.1) $&$2.5 \; (\pm 3.6) $ \\
 22&$\{12133\bar{1}\bar{2}\bar{1}\bar{3}\bar{3}\}$
    &$ 14.5 \; (\pm 4.4) $&$2.5 \; (\pm 3.7) $ \\
 23&$\{1234\bar{3}\bar{1}\bar{4}\bar{2}\}$
    &$ 12.6 \; (\pm 4.8) $&$2.4 \; (\pm 3.8) $ \\
 24&$\{12\bar{1}\bar{2}12\bar{1}\bar{2}\}$
    &$ 52.8 \; (\pm 7.8) $&$2.4 \; (\pm 3.8) $ \\
 25&$\{1234\bar{3}\bar{2}\bar{1}\bar{4}\}$
    &$ 12.6 \; (\pm 4.3) $&$2.4 \; (\pm 3.8) $ \\
 26&$\{12213\bar{1}\bar{2}\bar{2}\bar{1}\bar{3}\}$
    &$ 10.7 \; (\pm 4.4) $&$2.3 \; (\pm 3.7) $ \\
 27&$\{11122\bar{1}\bar{1}\bar{1}\bar{2}\bar{2}\}$
    &$ 32.6 \; (\pm 5.6) $&$2.3 \; (\pm 3.8) $ \\
 28&$\{1123\bar{1}\bar{1}\bar{2}\bar{3}\}$
    &$ 10.5 \; (\pm 4.7) $&$2.3 \; (\pm 3.7) $ \\
 29&$\{112334\bar{3}\bar{3}\bar{2}\bar{1}\bar{1}\bar{4}\}$
    &$ 12.3 \; (\pm 4.5) $&$2.3 \; (\pm 3.8) $ \\
 30&$\{122\bar{1}31\bar{2}\bar{2}\bar{1}\bar{3}\}$
    &$ 10.9 \; (\pm 4.5) $&$2.3 \; (\pm 3.8) $ \\
 31&$\{1123\bar{2}\bar{1}\bar{1}\bar{3}\}$
    &$ 10.9 \; (\pm 4.4) $&$2.3 \; (\pm 4.0) $ \\
 32&$\{11234\bar{3}\bar{2}\bar{1}\bar{1}\bar{4}\}$
    &$ 10.4 \; (\pm 4.4) $&$2.3 \; (\pm 4.0) $ \\
 33&$\{1122\bar{1}\bar{1}\bar{2}\bar{2}\}$
    &$ 26.6 \; (\pm 4.9) $&$2.2 \; (\pm 4.0) $ \\
 34&$\{12\bar{1}\bar{2}1\bar{2}\bar{1}2\}$
    &$ 71.4 \; (\pm 9.4) $&$2.2 \; (\pm 3.9) $ \\
 35&$\{12\bar{1}\bar{2}\}$
    &$ 22.6 \; (\pm 5.6) $&$2.2 \; (\pm 4.0) $ \\
\hline
   & $T$                                       &$295.0 \; (\pm 20.0)$  & \\
\end{tabular}
\end{table}


\newpage


\newpage


\newpage


\newpage


\newpage


\end{document}